\documentclass[preprint]{aastex6}

\usepackage{graphicx}
\usepackage{natbib}
\usepackage[english]{babel}  
\usepackage{subfigure}
\usepackage{url}
\newcommand{\new}{ }
\newcommand{\neww}{ }

\begin{document}

\title{Impulsive coronal heating from large-scale magnetic rearrangements: from IRIS to SDO/AIA}

%% Use \author, \affil, plus the \and command to format author and affiliation 
%% information.  If done correctly the peer review system will be able to
%% automatically put the author and affiliation information from the manuscript
%% and save the corresponding author the trouble of entering it by hand.
%%
%% The \affil should be used to document primary affiliations and the
%% \altaffil should be used for secondary affiliations, titles, or email.

%% Authors with the same affiliation can be grouped in a single
%% \author and \affil call.

\author{Fabio Reale\altaffilmark{1}}
\affiliation{Dipartimento di Fisica \& Chimica, Universit\`a di Palermo,
              Piazza del Parlamento 1, 90134 Palermo, Italy;
              fabio.reale@unipa.it}
\author{Paola Testa}
\affiliation{Center for Astrophysics | Harvard \& Smithsonian, 60 Garden St., Cambridge, MA 02138, USA}
\author{Antonino Petralia}
\affiliation{INAF-Osservatorio Astronomico di Palermo, Piazza del Parlamento 1, 90134 Palermo, Italy}
\author{David R. Graham}
\affiliation{Bay Area Environmental Research Institute, NASA Research Park, Moffett Field, CA , CA 94952, USA}

\altaffiltext{1}{and INAF-Osservatorio Astronomico di Palermo, Piazza del Parlamento 1, 90134 Palermo, Italy}

%% Mark off the abstract in the ``abstract'' environment. 
\begin{abstract}
The Interface Region Imaging Spectrograph (IRIS) has observed bright spots at the transition region footpoints associated with heating in the overlying loops, as observed by coronal imagers. Some of these brightenings show significant blueshifts in the Si\,{\sc iv} line at 1402.77~\AA\ ($\log T[K] \approx 4.9$). Such blueshifts \new{cannot} be reproduced by coronal loop models assuming heating by thermal conduction only, but \new{are} consistent with electron beam heating, highlighting for the first time the possible importance of non-thermal electrons in the heating of non-flaring active regions. Here we report on the coronal counterparts of these brightenings observed in the hot channels of the  Atmospheric Imaging Assembly (AIA) on board the Solar Dynamics Observatory. We show that the IRIS bright spots are the footpoints of very hot and transient coronal loops which clearly experience strong magnetic \new{interactions} and rearrangements, thus confirming the impulsive nature of the heating and providing important constraints for \new{their} physical interpretation. 
%We explore in detail how these hot loops might be produced through numerical 3D MHD modeling of interacting magnetic structures including the full plasma chromospheric and coronal response. 
\end{abstract}


\section{Introduction} 
\label{sec:intro}

Impulsive events play a major role in the corona. Flares are the most prominent but they are believed to scale down to a population of smaller events (nanoflares) \citep[e.g.,][]{Hudson1991a,Argiroffi2008a,Aschwanden2008c,Hannah2011a}. 

One key
question concerns details of the properties and origin of the impulsive energy
release. Signatures of the energy release itself are often obscured by the violent
heating and ionization in the lower atmosphere, sometimes saturating detectors \citep[e.g.,][]{Lin2001a}. Accurate, occasionally fortuitous, positioning of the
instrument and fast sampling are essential to glimpse pre-impulsive epoch
signatures \citep[see][]{Jeffrey2018a} although such observations are rare. At the other extreme, a diffuse nanoflaring activity is extremely elusive, because nanoflare storms driven by chaotic magnetic braiding \citep[e.g.,][]{Parker1988a} leave faint and ambiguous signatures \citep[e.g.,][]{Klimchuk2006a}. 

Events at intermediate scales are therefore fundamental to capture the basic mechanisms of impulsive energy releases in the solar corona. Events recently captured by the \new{Interface Region Imaging Spectrograph (IRIS) observations} are excellent candidates. As shown by \cite{Testa2014a}, \new{IRIS} detected significant Doppler shifts during the brightening of hot spots inside active regions observed in the transition region in the Si\,{\sc iv} line at 1402.77 \AA\ ($T \approx 10^{4.9}$~K).  The brightenings are highly variable, with a typical duration of 20 to 60 s.
Some of them show moderate blueshifts with typical velocities of $\sim 15$~km/s.
The blueshifts could not be reproduced by hydrodynamic models of plasma confined in a loop where the heat pulse is transported along the loop exclusively by thermal conduction. Instead, they were consistently reproduced by models where the heating is driven by beams of non-thermal electrons streaming down along the loop and hitting the dense plasma at the footpoints \citep{Testa2014a,Polito2018a}. The energy distribution of the electron beams is typically described with a power law, with a low energy cutoff $E_C$. Comparison with the models \new{shows} that the observed blueshifted transition region brightenings can be reproduced by heating deposited on small time scales ($\leq 30$s), characterized by total energy of $\leq 10^{25}$~erg, and with low energy cutoff $E_c \sim 10$~keV \citep{Testa2014a,Polito2018a}, lower than in major flares where larger energies are involved \citep[e.g.,][]{Hannah2011a}. Although less energetic, the presence of such electron beams is a major indication of magnetic reconnection \citep{Priest2000b,Cargill2015a}.

The present study is devoted to a systematic study of the coronal counterparts of the  brightenings observed with IRIS (Testa et al., in preparation). As mentioned above, \cite{Testa2014a} \new{have} already shown that there is a clear correspondence between IRIS brightenings and the ignition of loop systems, in particular visible in hot EUV channels of the Atmospheric Imaging Assembly on-board the Solar Dynamics Observatory. Here we will focus on this correspondence and analyse the features of the loop systems.

Coronal loops outside of flares but at temperatures above 5-6 MK have been extensively observed and studied in the past. Most studies focused on demonstrating that these coronal loops were really the site of very hot plasma, with the detection of emission from single very hot lines \citep{Ko2009a,Testa2012c}, of very hot components in broad-band spectra \citep{Miceli2012a}, hard X-rays \citep{McTiernan2009a,Ishikawa2017a,Marsh2018a}, and imaging from narrow-band EUV \citep{Reale2011a,Brosius2014a} and broad-band X-rays \citep{Porter1995a,Reale2009b}. Also emission measure reconstruction recovered small very hot components in active region loops \citep{Petralia2014a,Parenti2017a,Ishikawa2017a}.
These studies either analyzed active regions as a whole, or distinctly single loops.

In this study we will show that very hot loops corresponding to IRIS hot spots are to be treated as systems of interacting loops, and it is probably this interaction that determines the high temperature and the coherent behavior that make them intermediate between proper single flares and storms of nanoflares. 
%We are also developing a 3D MHD model of interacting loops (which will be presented in a follow-up paper, Petralia et al., in preparation), which is inspired by these observations, and aimed at providing a possible scenario to interpret the observed evolution of these nanoflare heated loops.

Section~\ref{sec:obs} describes the data,  and the results are discussed in Section~\ref{sec:discus}.

\section{The observations and data analysis} 
\label{sec:obs}

The Atmospheric Imaging Assembly \citep[AIA;][]{Lemen2012a} on board the Solar Dynamics Observatory (SDO) is based on normal-incidence optics and is equipped with six narrowband filters in the EUV band (94~\AA~ to 335~\AA) that contain a few strong spectral lines that sample the solar corona in a wide temperature range, approximately between 0.5 and 10~MK \citep{Boerner2012a,Boerner2014a}. The instrument continuously monitors the full-disk corona with high cadence (􏰊12~s) and a pixel size of $\sim 0.6$~arcsec. 

\begin{figure}              %%%%%%Figura%%%%%%%%%
 \centering
 \subfigure[]
   {\includegraphics[width=8.7cm]{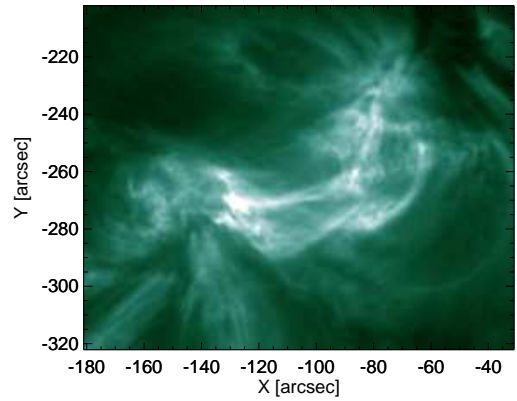}}
 \subfigure[]
   {\includegraphics[width=9.5cm]{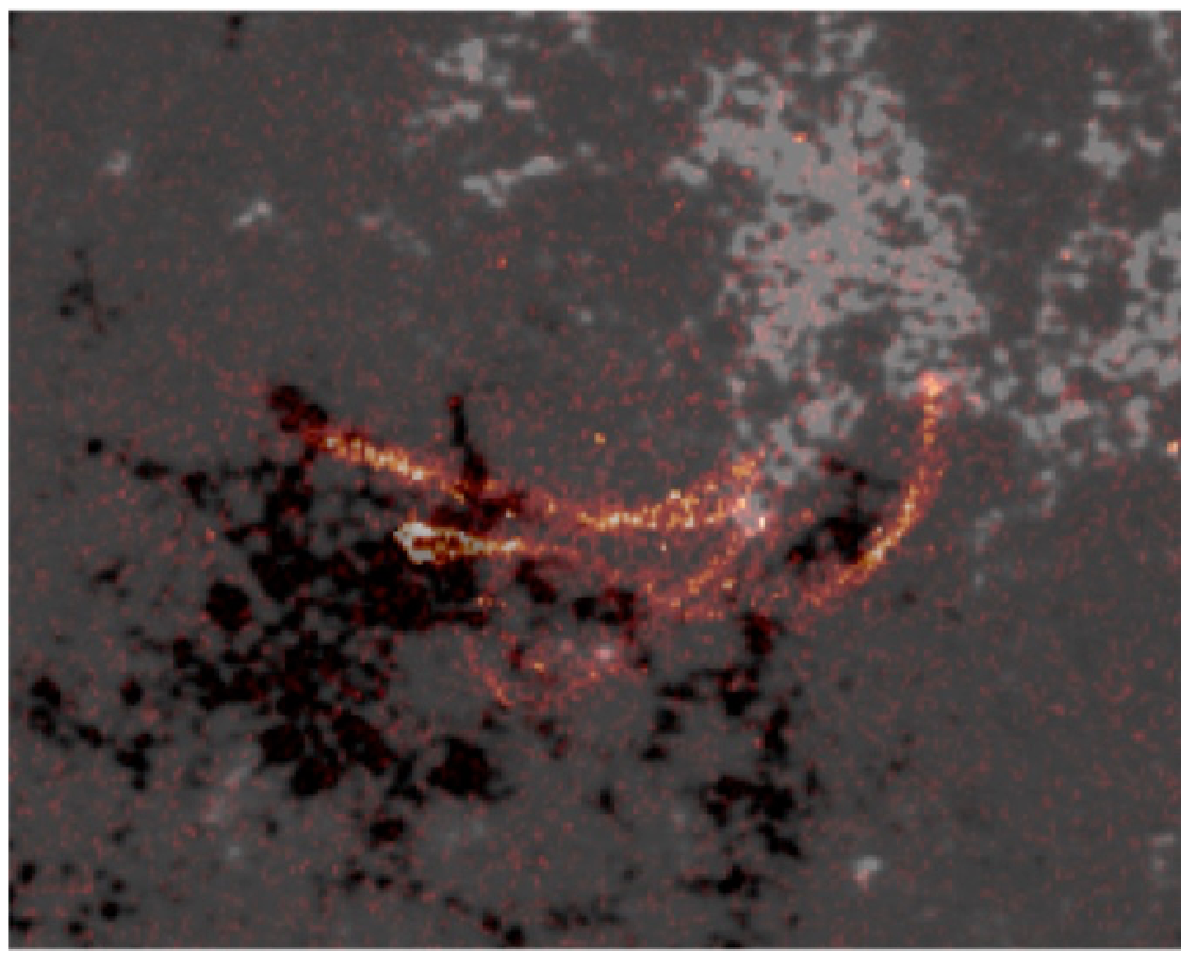}}
\subfigure[]
   {\includegraphics[width=8.7cm]{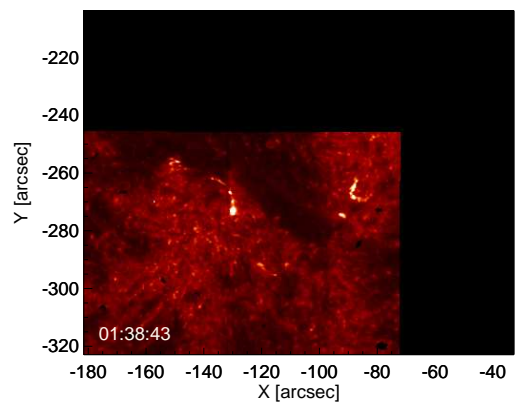}}
\caption{\footnotesize Loop system overlying short lived UV transition region brightenings detected by IRIS, observed on 12 november 2015 (Event 9 in Table~\ref{table1}). (a) Field of view as observed in the AIA 94~\AA\ channel after integrating 100 images (no background subtracted). (b) Magnetogram ($-100 < B < 100$~G, black and white) over 94~\AA\ image (red, panel c in Fig.~\ref{fig:reg1_94}) (c) \neww{IRIS Si\,{\sc iv} 1400~\AA\ passband} image at the labelled time. The coordinates are heliocentric.}
\label{fig:reg1_fov}
\end{figure}

In this study we focus on the evolution of hot arch-like structures whose footpoints correspond to hot spots detected \neww{in the Si\,{\sc iv} 1400~\AA\ passband with IRIS}. In particular, we analyze the evolution observed in the two AIA channels \new{that are} most sensitive to plasma at temperatures higher than 5~MK.
The 94~\AA\ and 131~\AA\ channels include highly ionized Fe lines, Fe\,{\sc xviii} and Fe\,{\sc xxi} line, respectively, that are sensitive to plasma at temperatures in the range 6-8~MK, and 9-12~MK, respectively. \new{We point out that both 94~\AA\ and 131~\AA\ channels include also other intense Fe lines (Fe\,{\sc ix}, Fe\,{\sc x} for 94~\AA\ and Fe\,{\sc viii} for 131~\AA) \citep{Foster2011a,Lemen2012a} which generally dominate the sensitivity of these two passbands at the lower temperatures \citep[0.5-1 MK, e.g.,][]{Martinez-Sykora2011a}, which we are not interested in here. It is worth noting though that the cool (transition region) contribution to these bands is quite limited in hot active region core loops studied here \citep[e.g.,][]{Testa2012a}.}

Based on IRIS evidence \neww{in the Si\,{\sc iv} 1400~\AA\ passband}, we have selected 10 events \new{occurred} between February 2014 and December 2015, listed in Table~\ref{table1}. In all of them, we see brightenings in both hot AIA channels. All of them are below the threshold for flare detection in GOES light curves.

\begin{table}
\begin{center}
  \caption{Transient EUV events with corresponding hot spots in IRIS observations. The time (UT) is taken just before the beginning of the event, the one of the image for background subtraction (94~\AA). The coordinates [X,Y] are of the center of the dataset (arcsec).}
  \label{table1}
  \begin{tabular}{lcccc}
  \hline
N&Date&Time&X&Y \\
\hline
 1&2014-02-04&13:34:49&       399.7&   -77.6\\
 2&2014-02-23&23:24:37&169.4&-60.7\\
 3&2014-03-19&15:15:25&      71.2&     303.6\\
 4&2014-04-10&02:42:01&       822.7&   -134.3\\
 5&2014-09-17&14:52:01&  -138.1&     91.8\\
 6&2014-09-17&17:15:01&      -99.1&       74.1\\
 7&2014-09-18&08:06:01&       34.4&   74.7\\
 8&2015-01-29&18:29:01&       149.8&      -78.6\\
 9&2015-11-12&01:37:12&      -117.2 &      -329.6\\
10&2015-12-24&15:17:00&-597.7&-352.1\\
\hline

\end{tabular}
\end{center}
\end{table}

%[Events time range and center coordinates]

We have ascertained that all of the events share a set of common features that we
describe in the following. Furthermore, we show a more detailed analysis for 3 of
them where the evolution is particularly clear and \new{yields} good counts; one of them
we study in greater detail. 
%We have extracted data frames in the relevant time ranges and with sides between 256 and 512 pixels, i.e., between 154" and 307". 

The relevant image sequences have been preprocessed with the standard AIA software procedure and co-aligned.
In order to study the evolution of the coronal structures in greater detail, we have devised procedures to highlight them above the underlying present emission. Our approach to this has been to subtract the emission just before the brightenings. \new{This allows us to subtract the contribution from the plasma in the temperature range 0.5-1~MK and to study exclusively the transient brightening of the hotter plasma ($> 6$~MK).}

\begin{figure}              %%%%%%Figura%%%%%%%%%
 \centering
    \subfigure[]
   {\includegraphics[width=8cm]{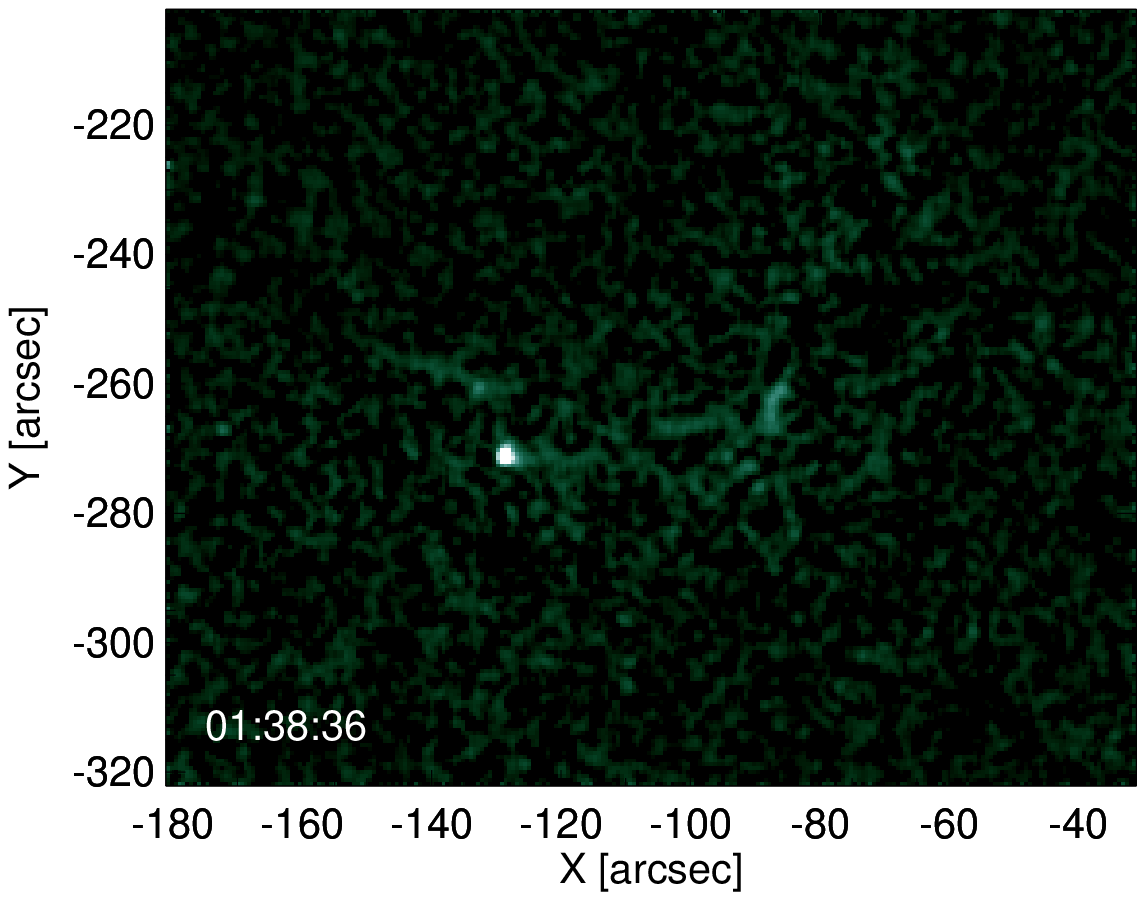}}
    \subfigure[]
   {\includegraphics[width=8cm]{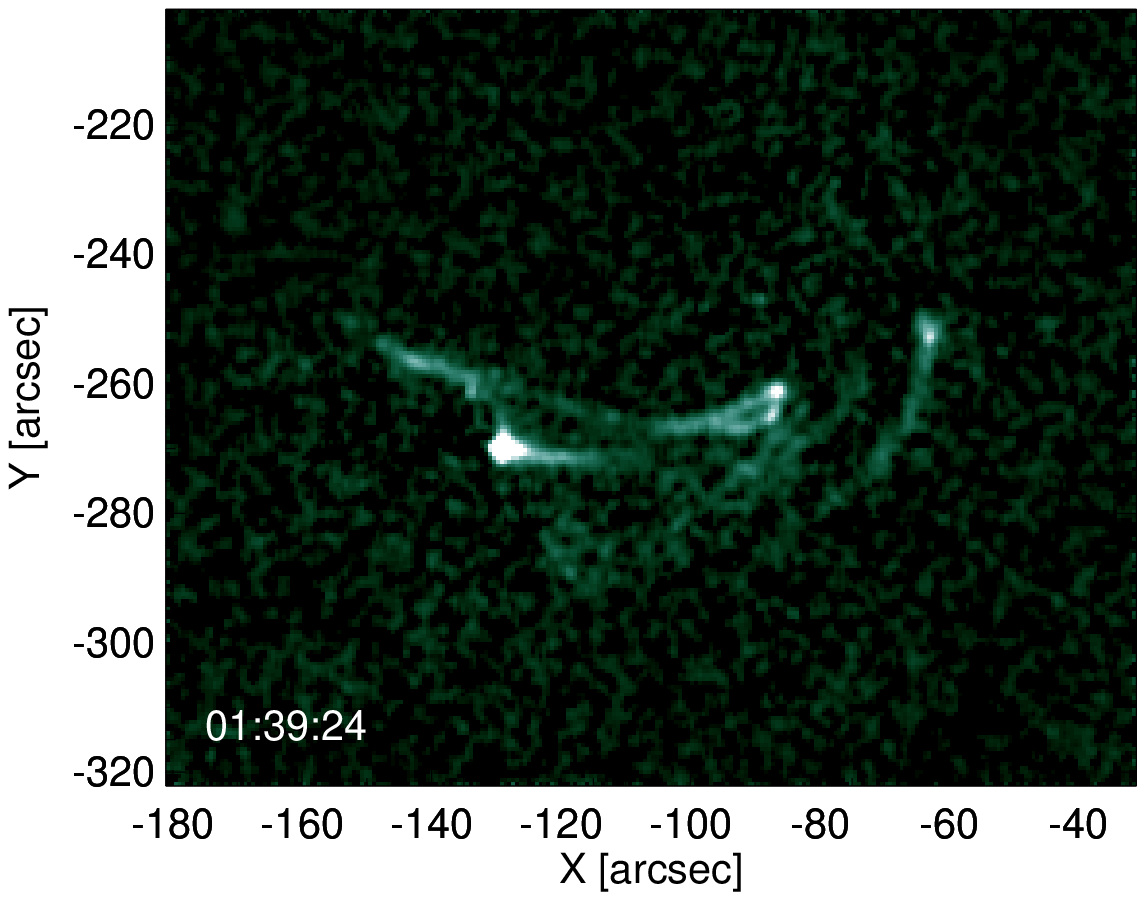}}
       \subfigure[]
   {\includegraphics[width=8cm]{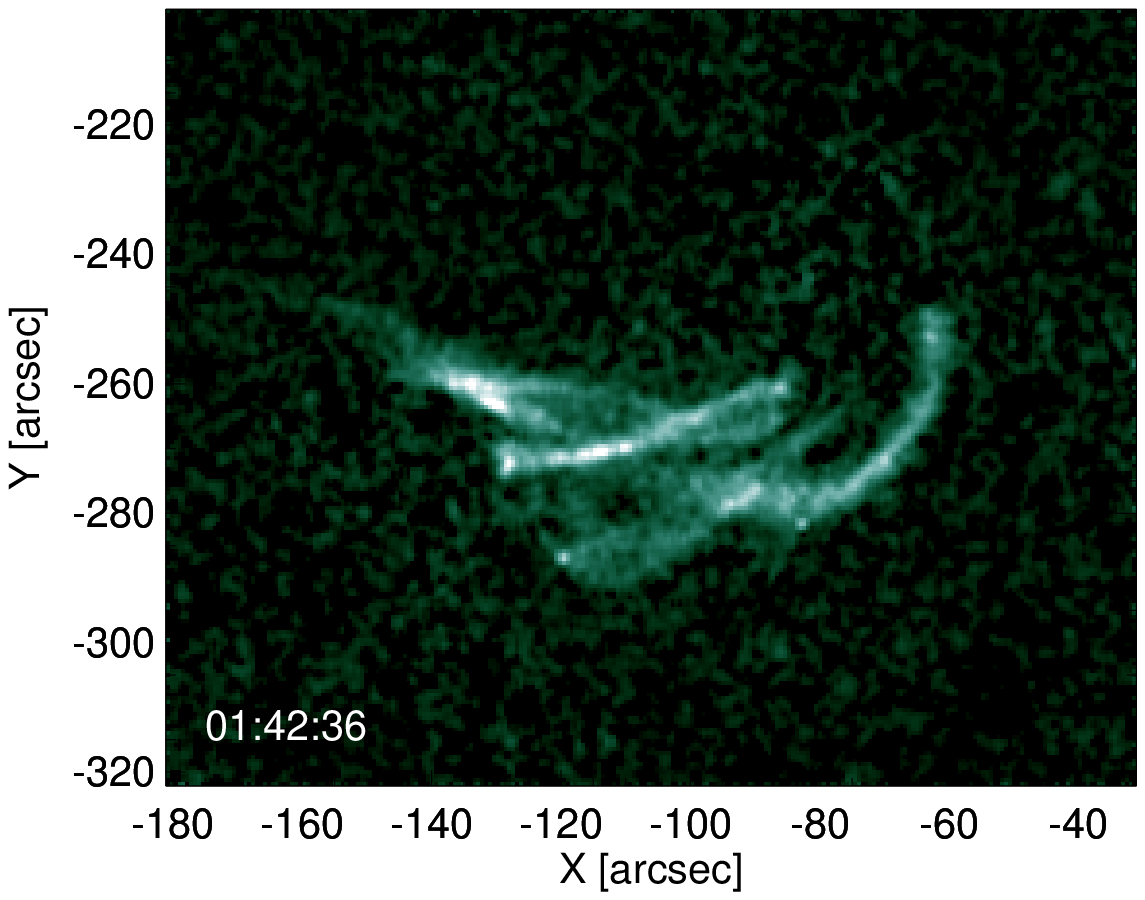}}
       \subfigure[]
   {\includegraphics[width=8cm]{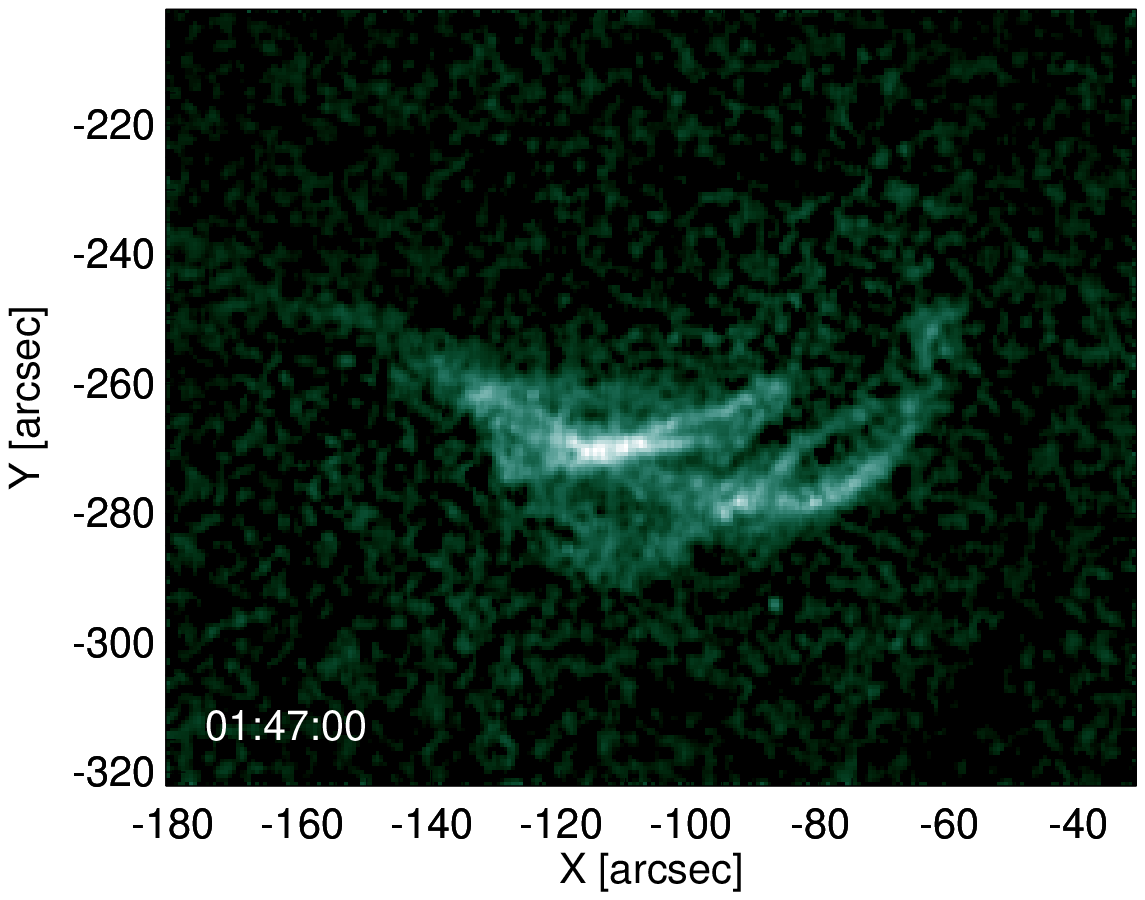}}       
   \subfigure[]
   {\includegraphics[width=8cm]{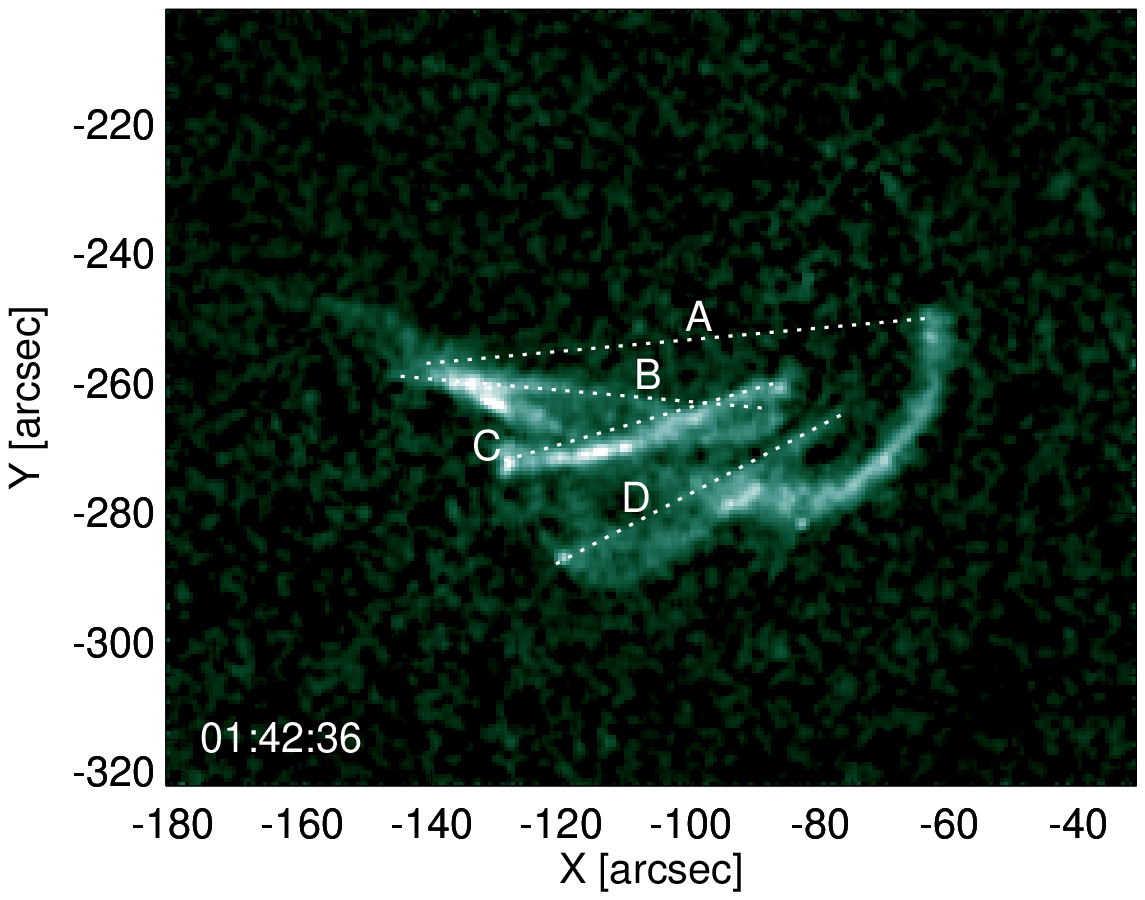}}       
   \subfigure[]
   {\includegraphics[width=8cm]{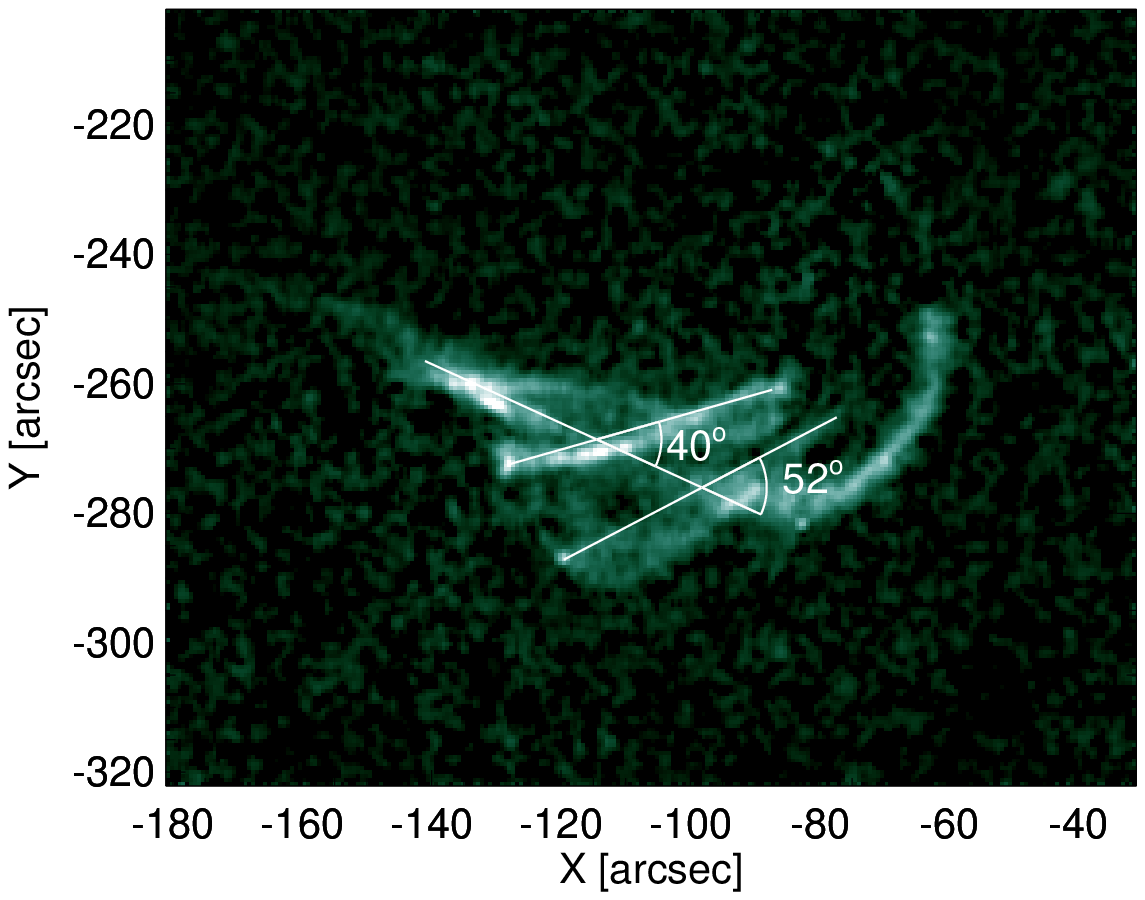}}
\caption{\footnotesize Loop system in Fig.~\ref{fig:reg1_fov}. (a)-(d) Evolution of the loop system in the 94~\AA\ channel after subtracting an image just before the brightening begins (1:37:12 UT). The frame times are labelled (see Movie~1, left-hand side, for an animation of this evolution); (e) relevant footpoint connections, (f) angles between intersecting arches.} 
\label{fig:reg1_94}
\end{figure}

\subsection{Loop evolution}

\begin{figure}              %%%%%%Figura%%%%%%%%%
 \centering
 \subfigure[]
   {\includegraphics[width=7cm]{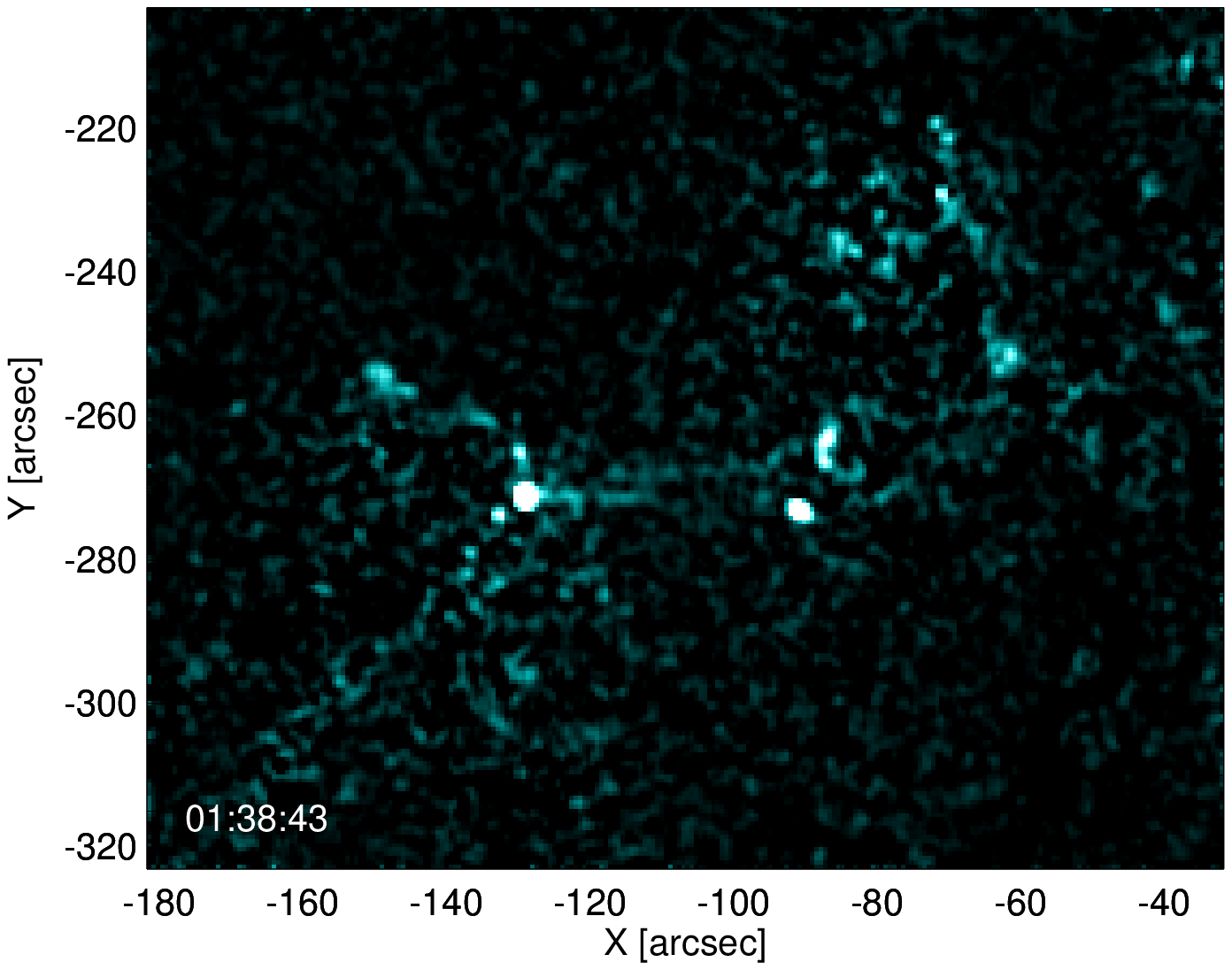}}
 \subfigure[]
   {\includegraphics[width=7cm]{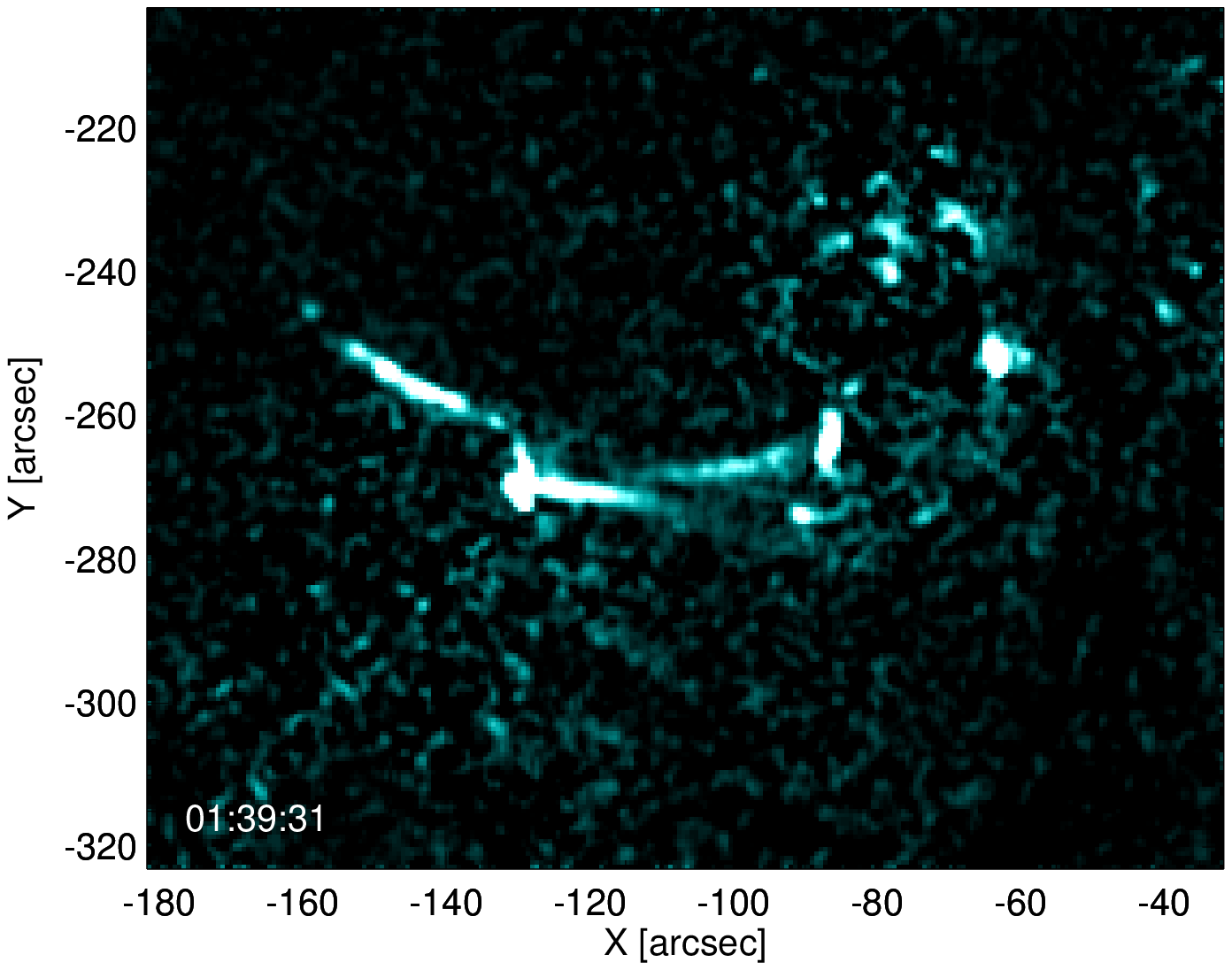}}
\subfigure[]
   {\includegraphics[width=7cm]{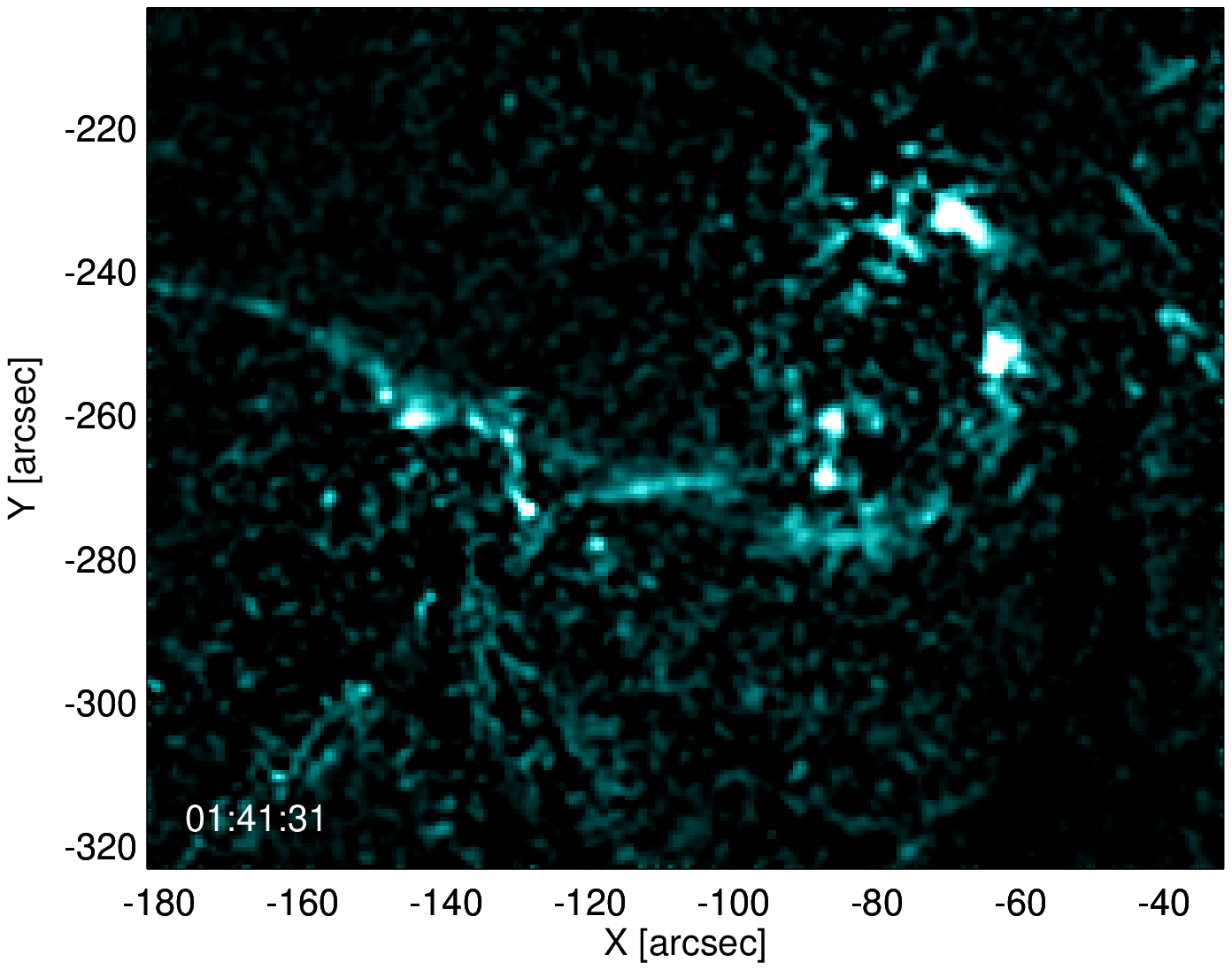}}
\caption{\footnotesize Three representative snapshots of the loop system of Fig.~\ref{fig:reg1_94} as observed in the AIA 131~\AA\ channel (see Movie~1, right-hand side, for the related animation).} 
\label{fig:reg1_131}
\end{figure}

We focus our attention on the event on 12 November 2015 (event 9 in Table~\ref{table1}) starting approximately at 1:38 UT.
Figures~\ref{fig:reg1_fov}-\ref{fig:reg1_131} (and Movie~1) show the loops system and its evolution. Figure~\ref{fig:reg1_fov} shows the \new{loop region} in the SDO/AIA 94~\AA\ channel, the corresponding magnetogram, and the same region as imaged with \neww{IRIS in the Si\,{\sc iv} 1400~\AA\ passband}. In panel~(a) the image is obtained after integrating over 100~images around the event. The system consists of several arches that \new{apparently intersect each other} at large angles ($\gtrsim 20^o$) \new{in the plane of the sky}. Moss-like structures are widespread in the image, which are related to the cool-sensitive side of the double-peaked channel response function. 

In panel (b) the magnetogram shows that the visible arches connect regions with opposite magnetic polarity. \new{The IRIS observation shows the bright features at some loop footpoints}. Figure~\ref{fig:reg1_94}a-d (and Movie~1, left side) shows the evolution of the loop system as imaged in the 94~\AA\ channel. The brightening clearly starts from the footpoints (Fig.~\ref{fig:reg1_94}a). Then the brightening propagates along the loop legs. At least three different loops appear to brighten almost at the same time (Fig.~\ref{fig:reg1_94}b). A couple of them \new{appears} to intersect along the line of sight (X=80", Y=60"). 
Around 4~minutes after the first brightening, a whole complex loop system appears to be bright (Fig.~\ref{fig:reg1_94}c): we distinguish several loops, a longer one that extends from left to right and is curved downwards. The system then begins to fade (Fig.~\ref{fig:reg1_94}d). The distance between the footpoints (see line~A in Fig.~\ref{fig:reg1_94}e) is approximately 57~Mm, which leads to a presumable upper limit for the total length of a semicircular loop of $\sim 90$~Mm. Among the others loops, we see at least two of them crossing the major one, whose distance between the footpoints are named C and D in Fig.~\ref{fig:reg1_94}e, and is 32~Mm and 36~Mm, respectively (upper limit for length 50~Mm and 57~Mm, respectively). As shown in Fig.~\ref{fig:reg1_94}f, the arc-like structures make an apparent angle of $40^o$ and $52^o$, respectively, \new{in the plane of the sky}.

Figure~\ref{fig:reg1_131} shows three snapshots of the loop system taken in the \new{131~\AA\ } channel after background subtraction (see Movie~1, right side, for the evolution in this channel). In general, in this channel we see fewer bright loops at a time and the 131 emission is shorter lived, compared to the cooler 94~\AA\ emission. In the earlier image, the two footpoints that brighten first are clearly visible. In the second image, we see a very similar topology as in the corresponding image in the 94~\AA\ channel (Fig.~\ref{fig:reg1_94}b). In the third one, we can distinguish the longest loop structure.

\begin{figure}              %%%%%%Figura%%%%%%%%%
 \centering
   {\includegraphics[width=8cm]{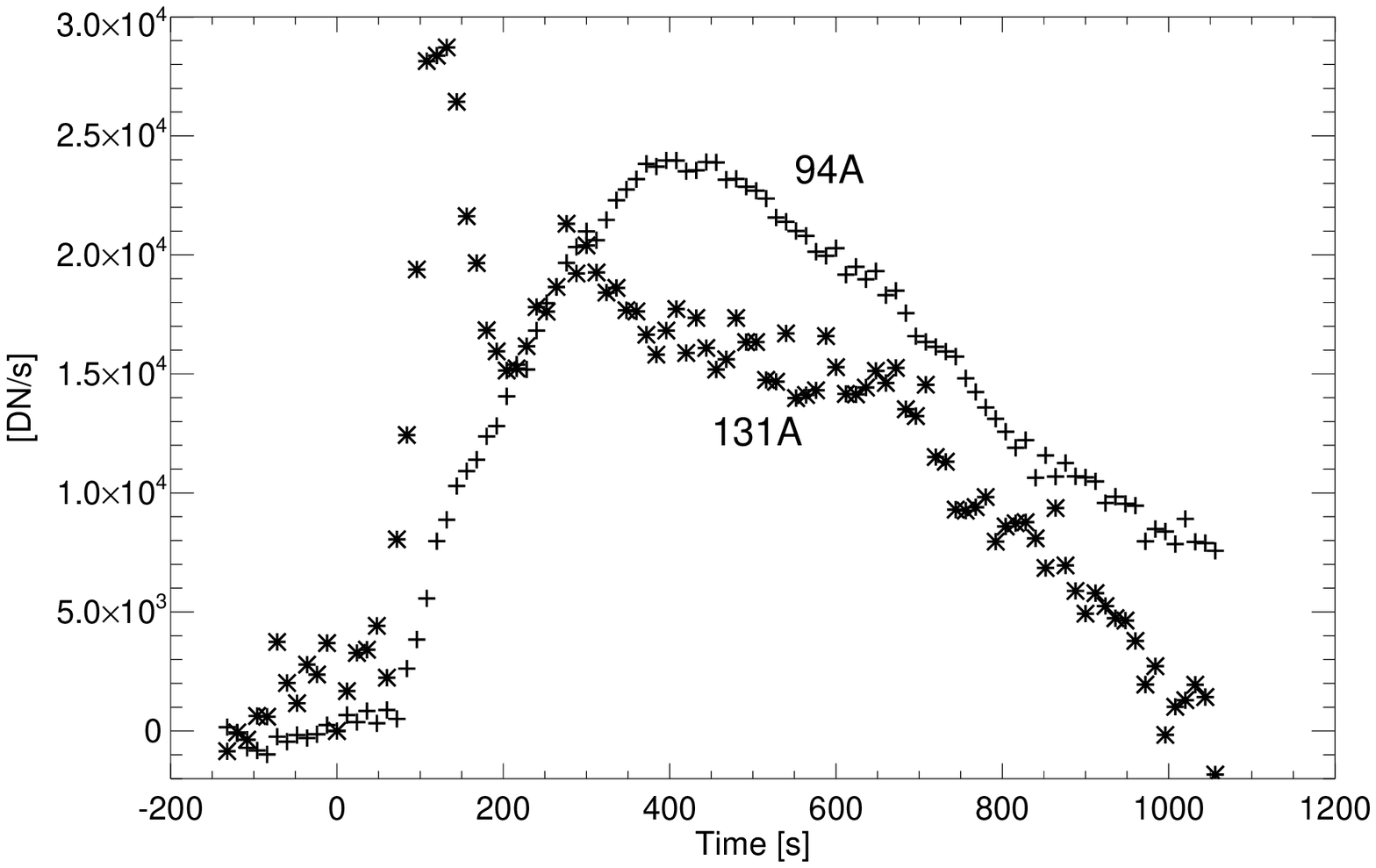}}
\caption{\footnotesize Light curves integrated over the frames in Fig.~\ref{fig:reg1_94}-\ref{fig:reg1_131} in the 94~\AA\ (crosses) and 131~\AA\ (stars) channels (seconds since 01:37:12 UT). The emission is background subtracted.} 
\label{fig:reg1_lc}
\end{figure}

This evolution has a clear correspondence in the light curves integrated \new{over the frames in Fig.~\ref{fig:reg1_94}-\ref{fig:reg1_131}, i.e., over the whole loop system,} in both channels, shown in Fig.~\ref{fig:reg1_lc}. In the 131~\AA\ channel, the emission has an early sharp peak (time $t \approx 400$~s), it remains relatively steady for the next $\sim 10$~min, and then decays rapidly. In the 94~\AA\ channel, the evolution is more gradual and delayed. The emission smoothly reaches its peak at $t \approx 650$~s, i.e., $\sim 4$~min later than in the other channel, and then continuously and gradually decreases.

\begin{figure}              %%%%%%Figura%%%%%%%%%
 \centering
 \subfigure[]
   {\includegraphics[width=8cm]{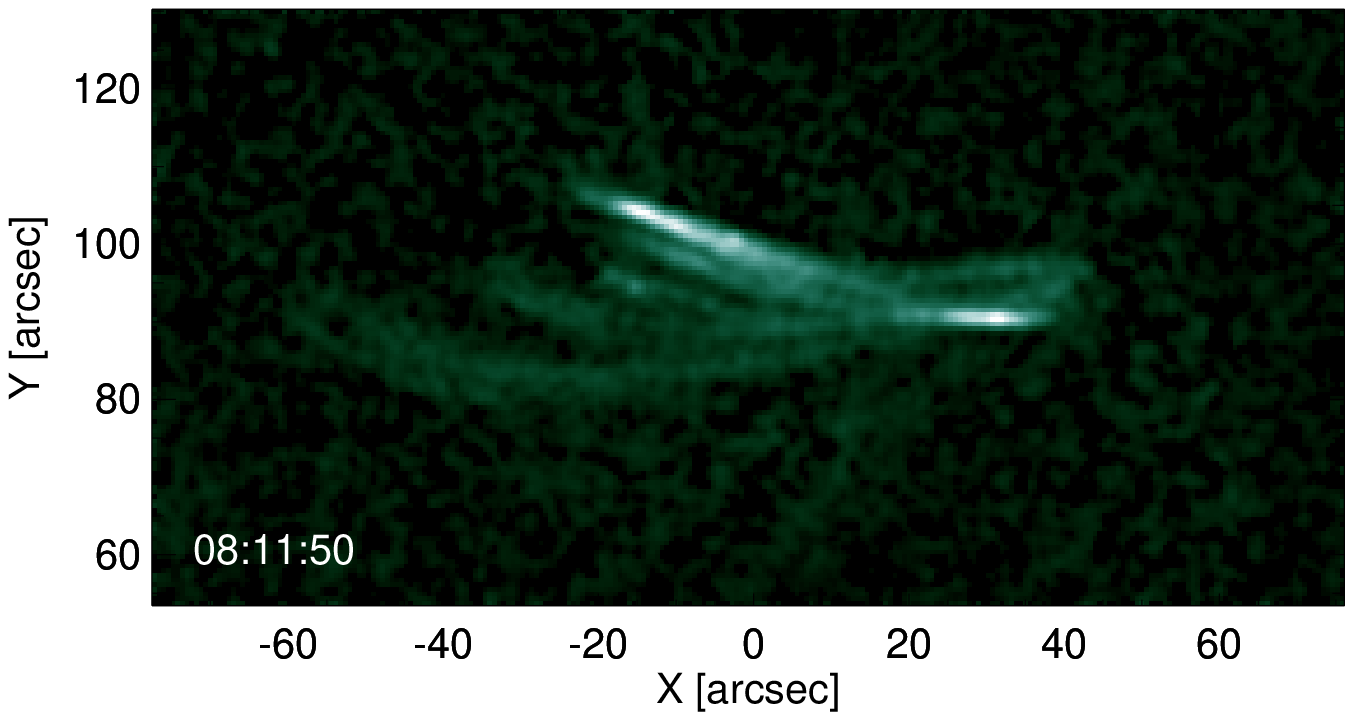}}
 \subfigure[]
   {\includegraphics[width=8cm]{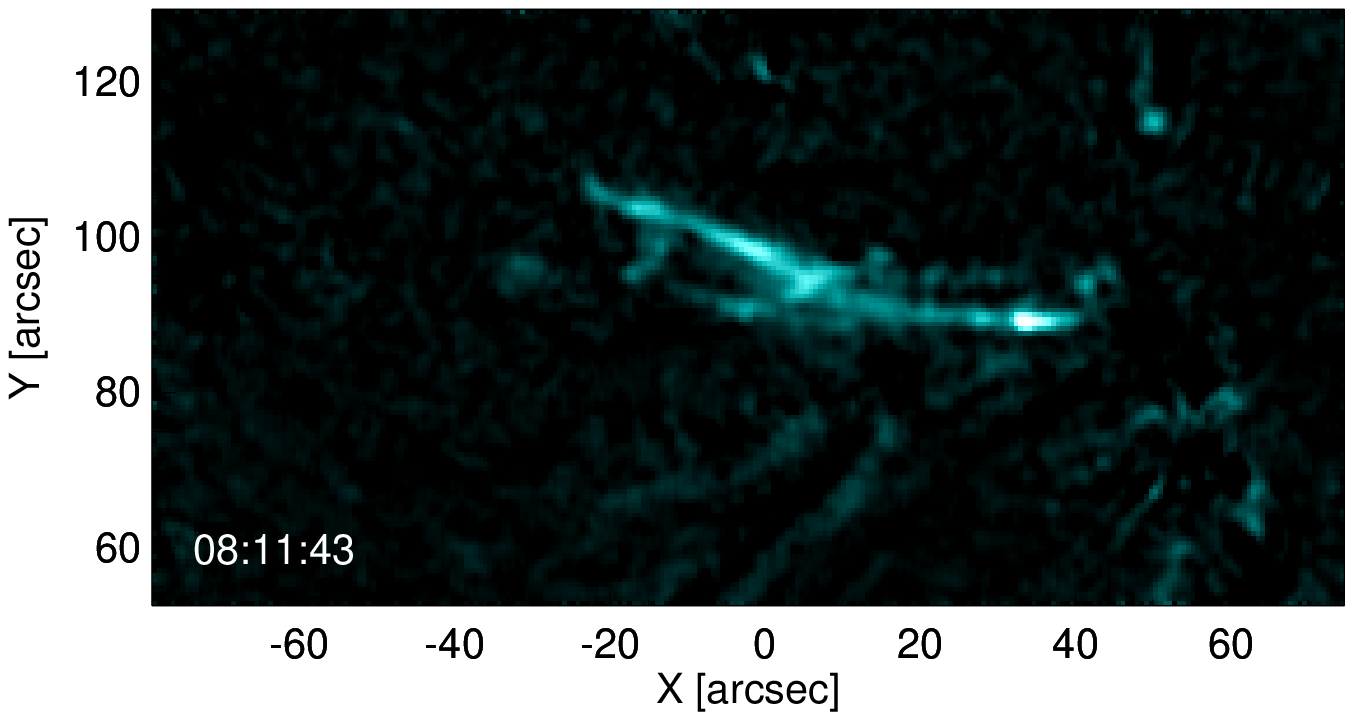}}
    \subfigure[]
   {\includegraphics[width=8cm]{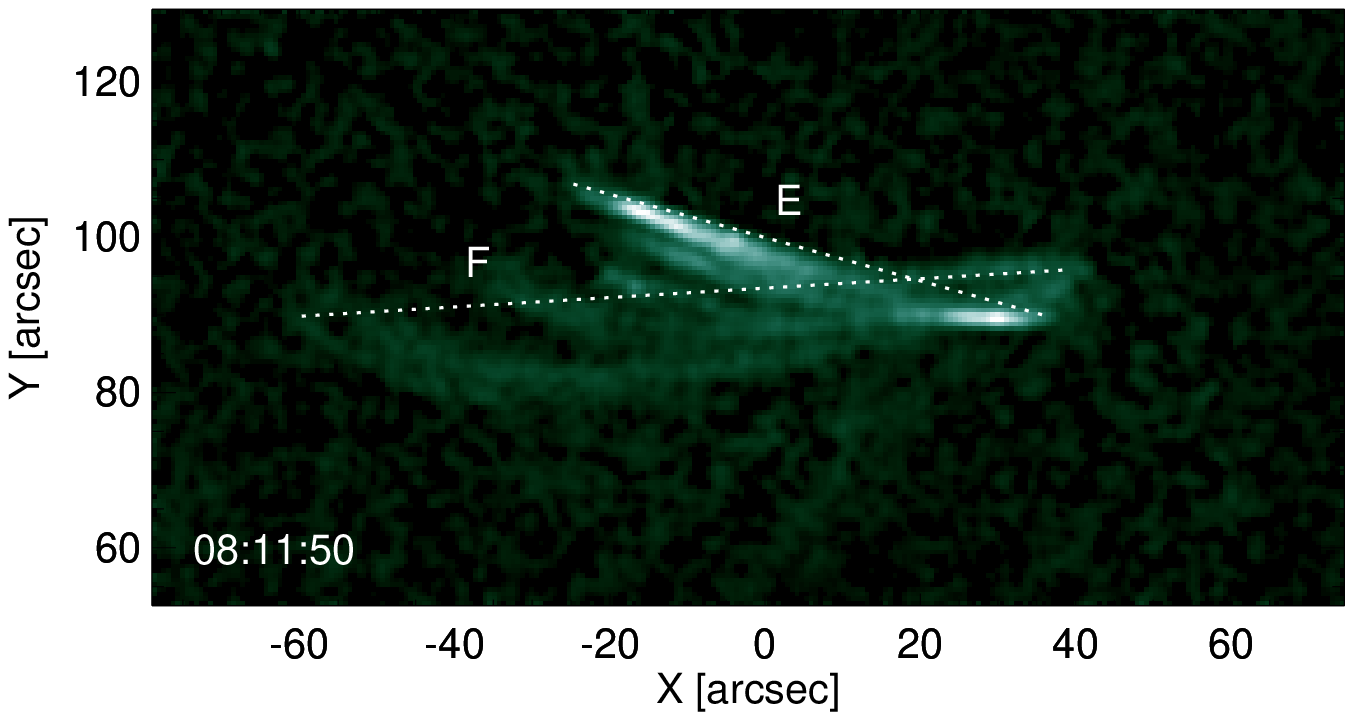}}
    \subfigure[]
   {\includegraphics[width=8cm]{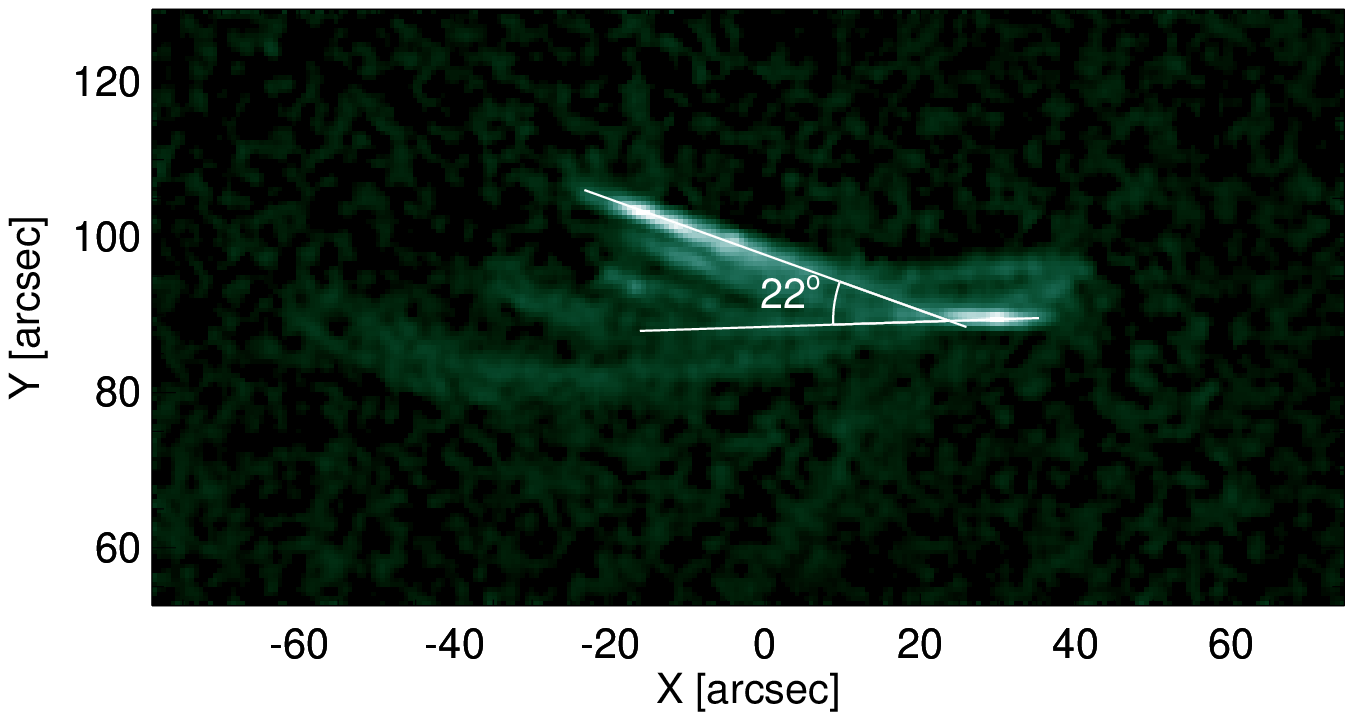}}
    \subfigure[]
   {\includegraphics[width=8cm]{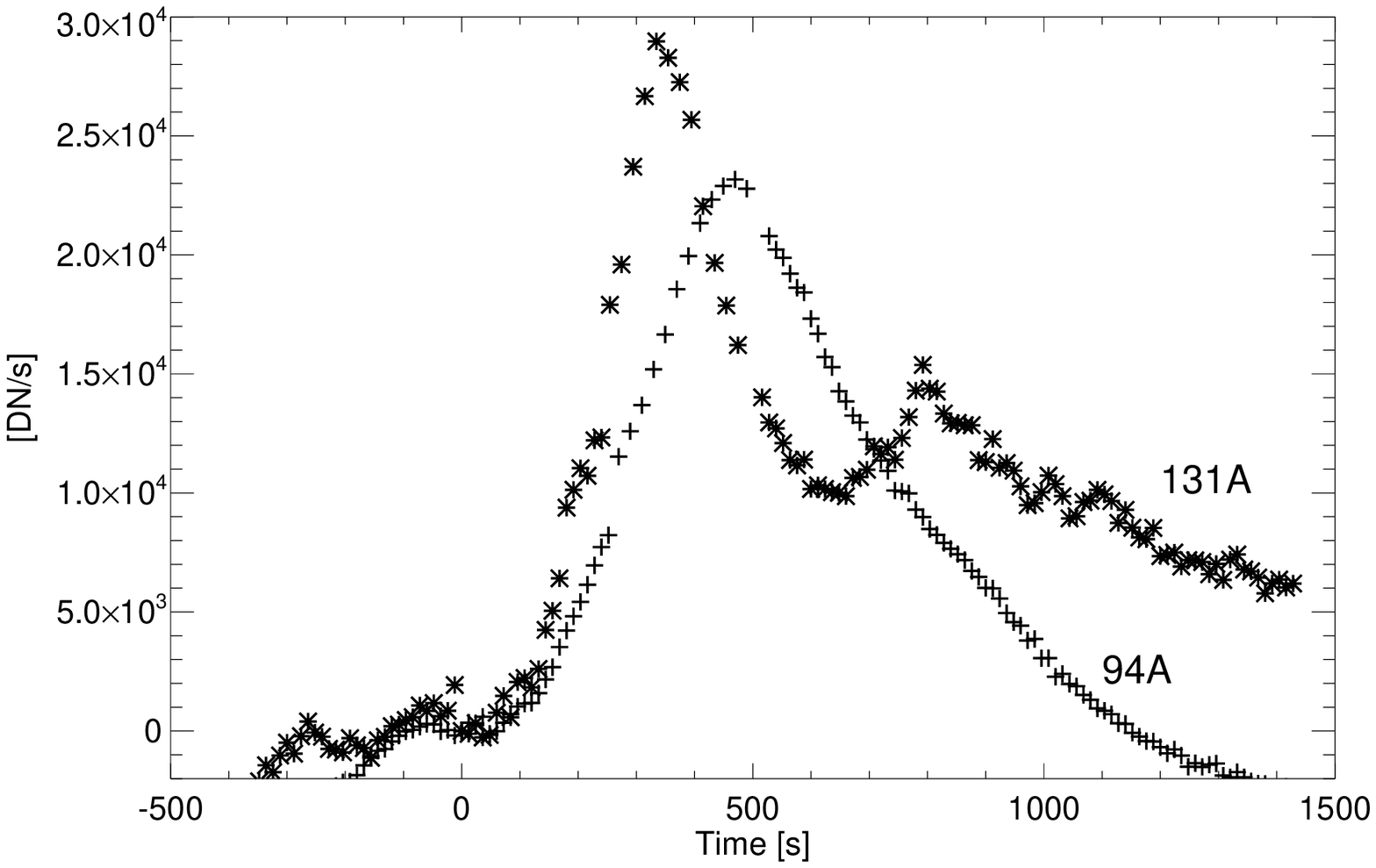}}
\caption{\footnotesize Loop system that corresponds to UV bright spots detected by IRIS, observed on 18 september 2014 (Event 7 in Table~\ref{table1}). Background subtracted image of the loop system at its maximum brightness observed (a) in the AIA 94~\AA\ channel and (b) in the 131~\AA\ channel (see Movie~2 for the related animation); (c,d) As Fig.~\ref{fig:reg1_94}e,f; (e) Light curves (seconds since 08:06 UT).} 
\label{fig:reg2}
\end{figure}

\begin{figure}              %%%%%%Figura%%%%%%%%%
 \centering
 \subfigure[]
   {\includegraphics[width=8cm]{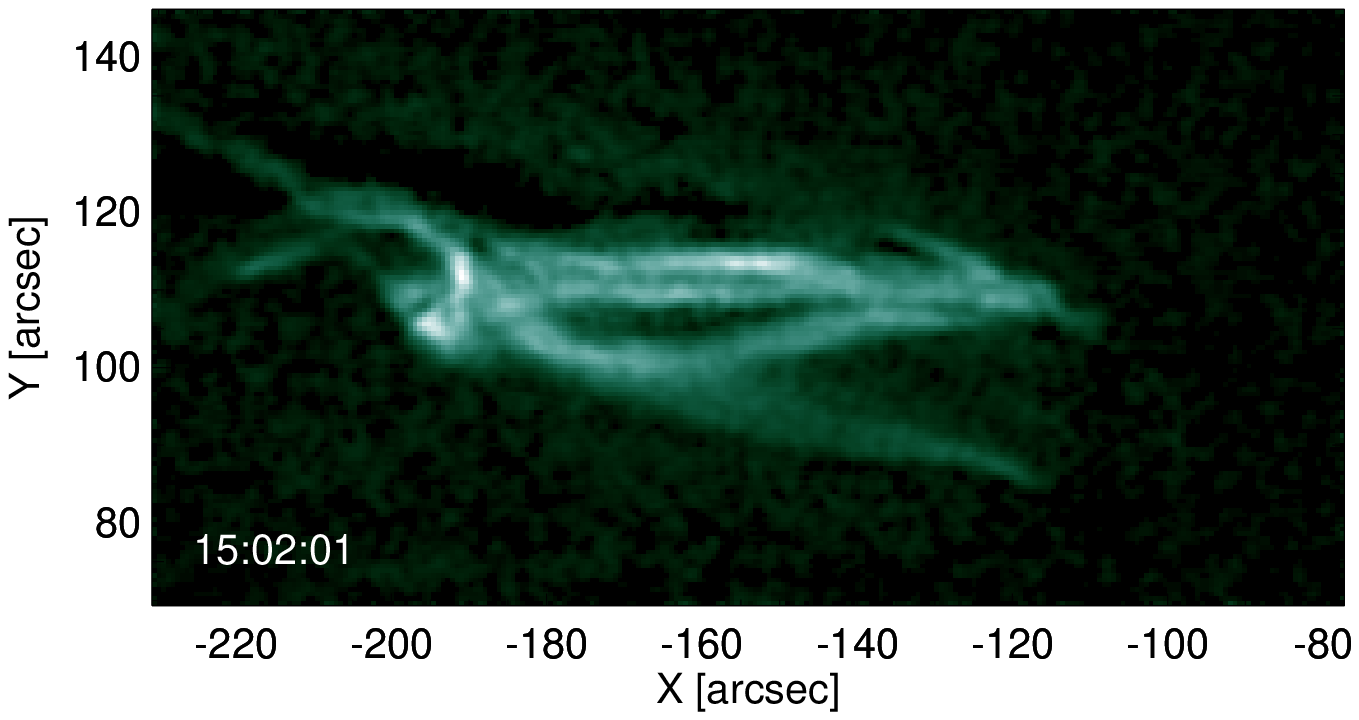}}
 \subfigure[]
   {\includegraphics[width=8cm]{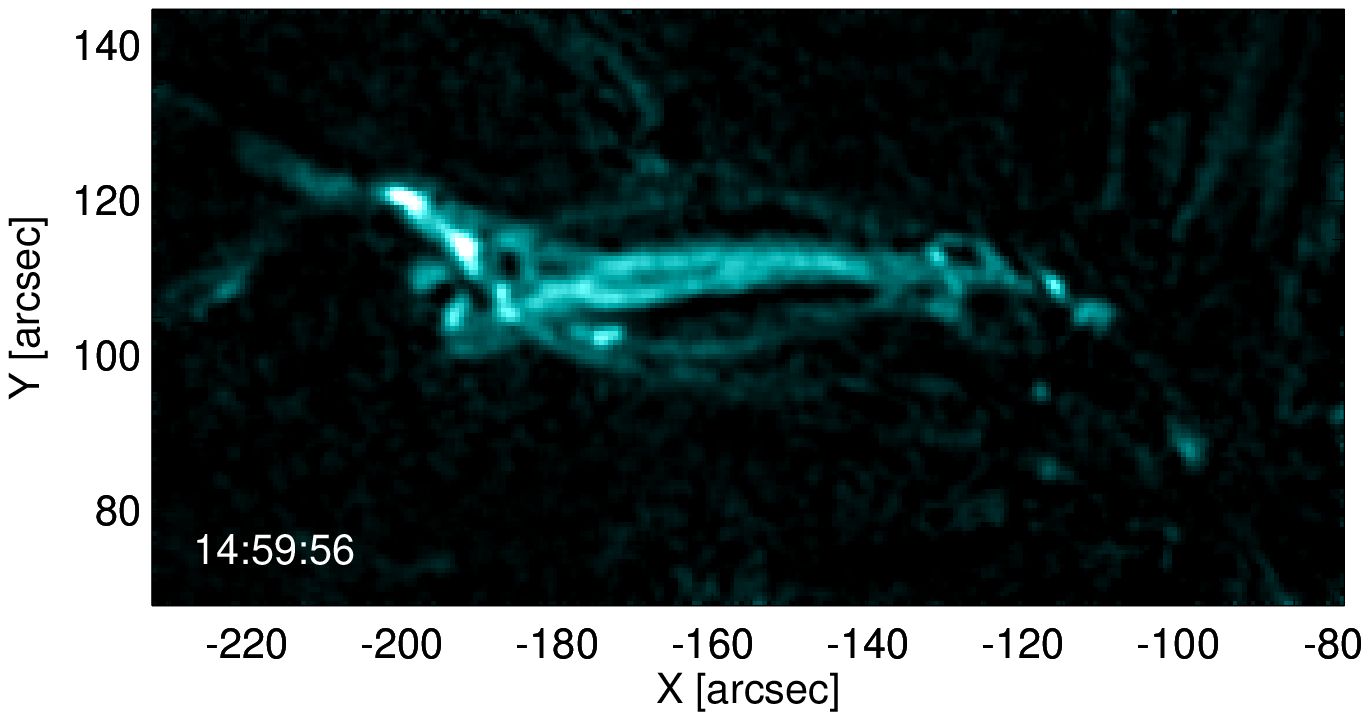}}
    \subfigure[]
   {\includegraphics[width=8cm]{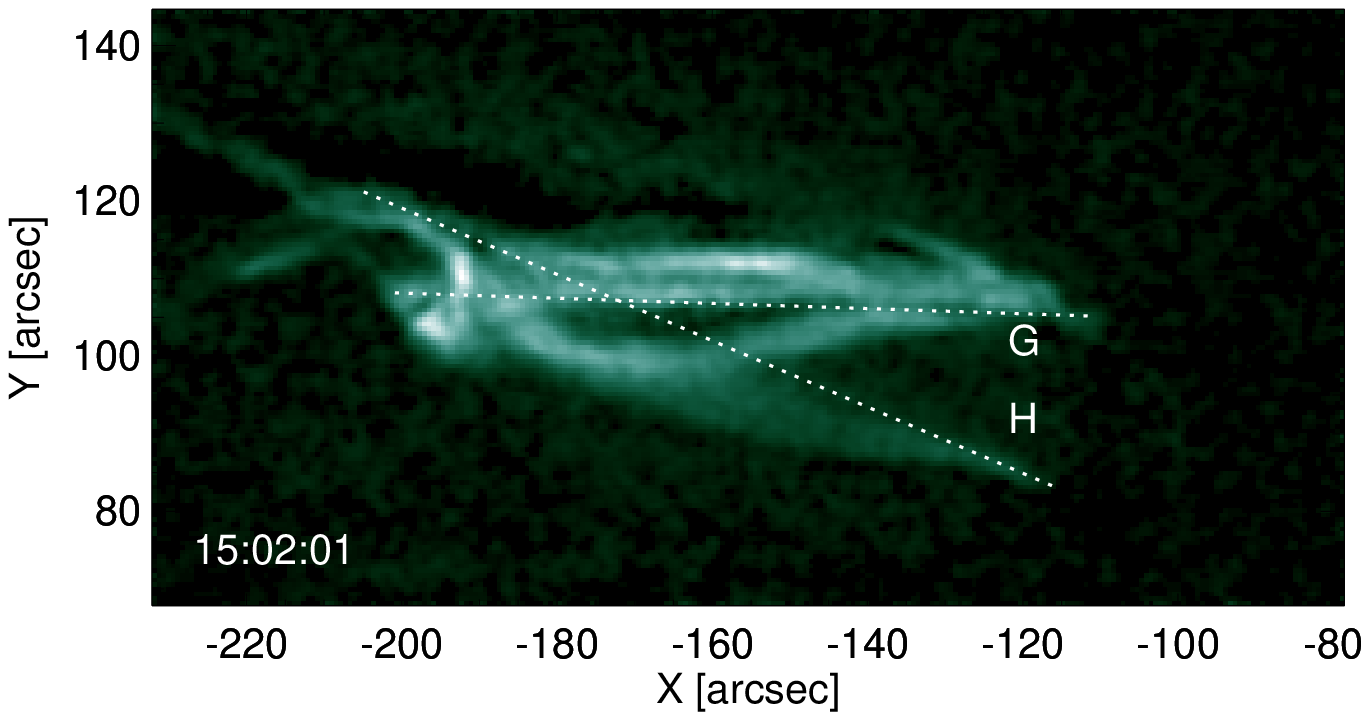}}
    \subfigure[]
   {\includegraphics[width=8cm]{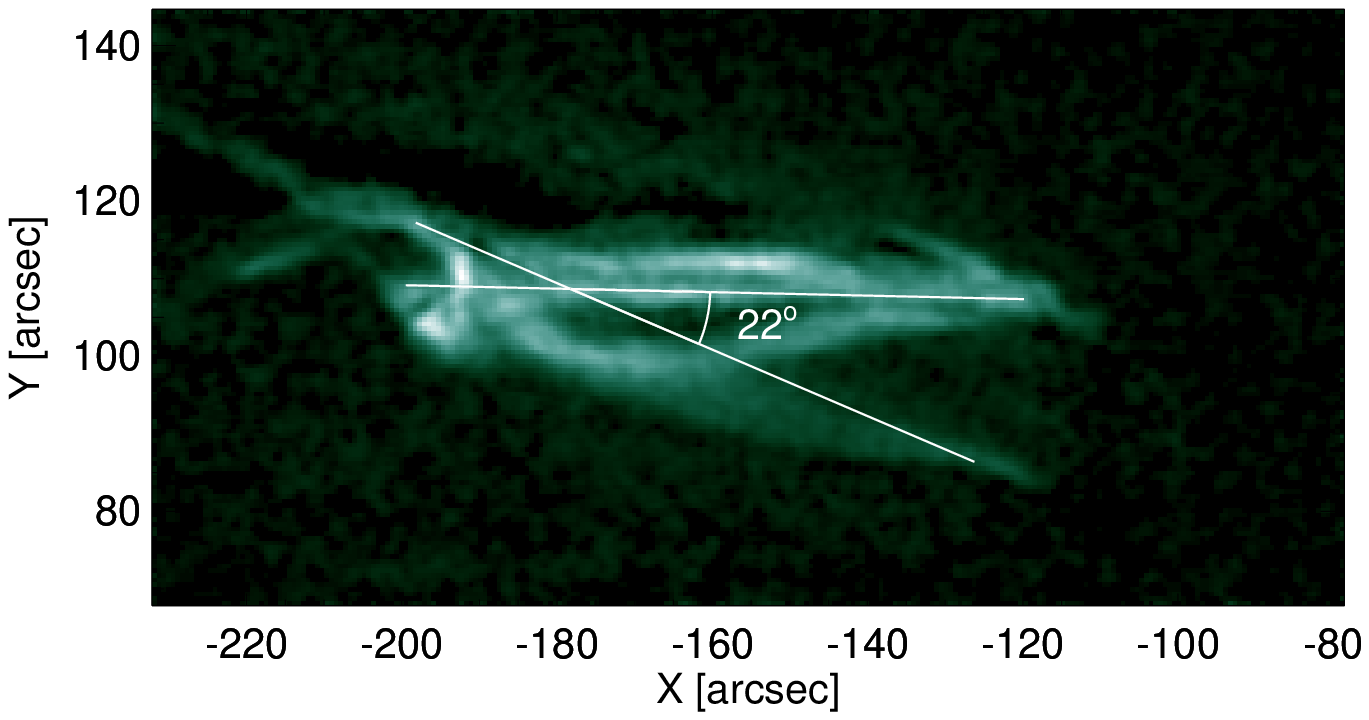}}
    \subfigure[]
   {\includegraphics[width=8cm]{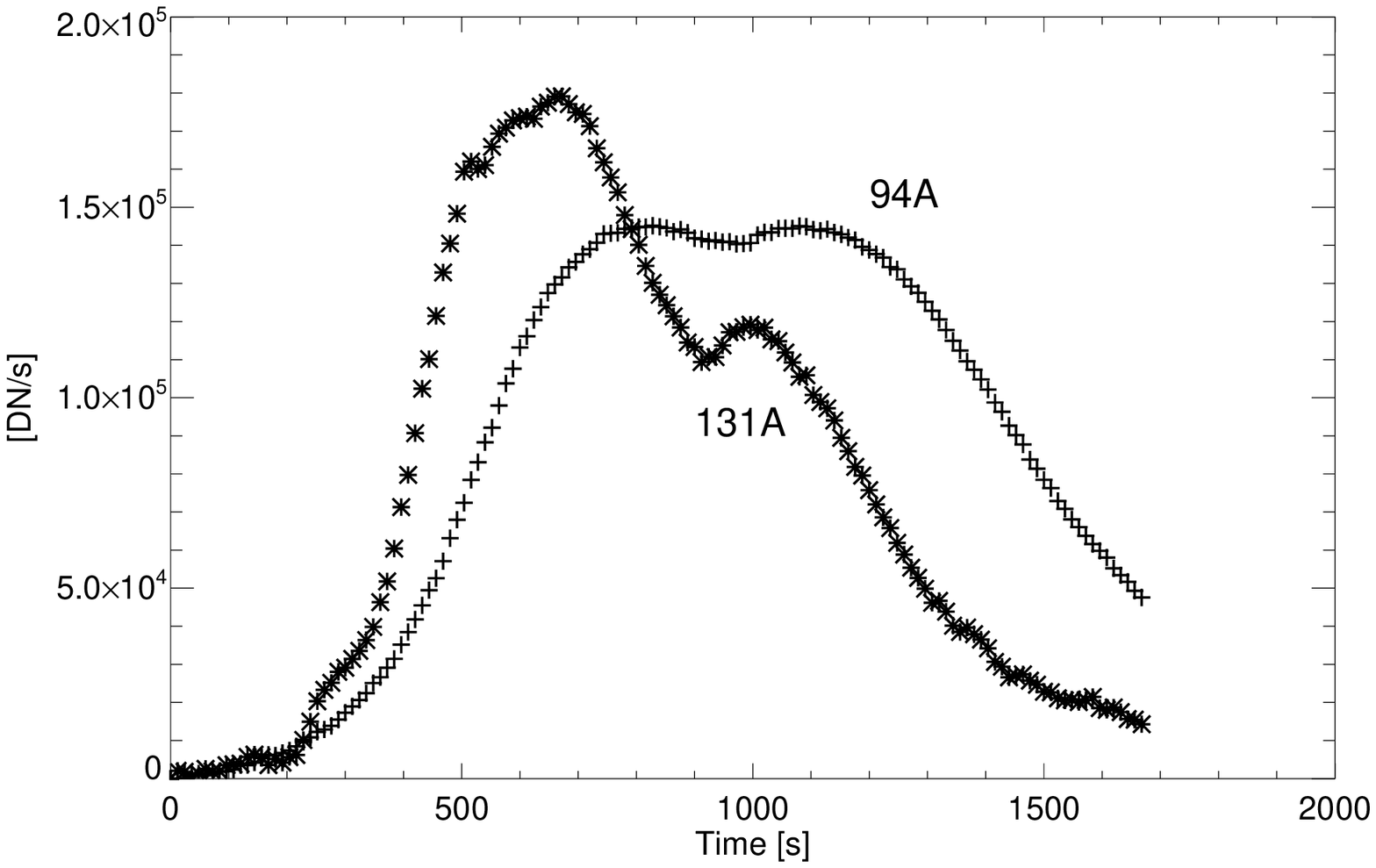}}
\caption{\footnotesize As in Fig.~\ref{fig:reg2} for the loop system that corresponds to UV bright spots detected by IRIS, observed on 17 september 2014 (Event 5 in Table~\ref{table1}, seconds since 14:52:01~UT, see Movie~3 for the related animation). } 
\label{fig:reg3}
\end{figure}

A couple of other cases are shown in Figs.~\ref{fig:reg2}-\ref{fig:reg3} (events 7 and 5 in Table~\ref{table1}, respectively). The loop system shown in Fig.~\ref{fig:reg2} is simpler than that in Fig.~\ref{fig:reg1_94} but similarly shows the presence of a mis-aligned loop bundle (Fig.~\ref{fig:reg2}a). Entangled loops are visible in the 131~\AA\ channel (Fig.~\ref{fig:reg2}b). The overall evolution is faster and the light curves simpler too, but they share with the previous case the earlier 131~\AA\ peak and the smoother curve in the 94~\AA\ channel (Fig.~\ref{fig:reg2}e). The footpoint distances marked in Fig.~\ref{fig:reg2}c are 46~Mm (E) and 72~Mm (F) \new{(upper limit for loop length 72~Mm and 113~Mm, respectively)}. A representative angle between misaligned structures is $22^o$ as shown in Fig.~\ref{fig:reg2}d.

The loop system shown in Fig.~\ref{fig:reg3} is more complex than that in Fig.~\ref{fig:reg2}. As clear in the 94~\AA\ channel (Fig.~\ref{fig:reg3}a), there are both quasi-parallel, probably entangled, loops and other oblique but intersecting loops. The whole system appears as a more compact bundle than the other two. The core loops are very bright also in the 131~\AA\ channel (Fig.~\ref{fig:reg3}b). As representative size, we measure footpoint distances of 65~Mm (G) and 69~Mm (H) (Fig.~\ref{fig:reg3}c) and a representative angle between misaligned structure is $22^o$ (Fig.~\ref{fig:reg3}d). Also in this case \new{the rise and the peak in the light curve in the 131~\AA\ channel are anticipated with respect to those} in the 94~\AA\ channel (Fig.~\ref{fig:reg3}e). However, at variance from the other cases, here we clearly see two different peaks in both channels, one $\sim 5$~min after the other, which most probably mark the presence of two distinct heating episodes.

\subsection{Emission measure evolution}

One key question of this analysis is whether the plasma that becomes transiently bright in the 94~\AA\ and 131~\AA\ channels is really very hot. Although the simultaneous brightening in the two channels after background subtraction and their timing is perfectly compatible with them heated to and cooling from temperatures above $10^7$~K, we derive more constraints about the thermal composition of the brightening plasma from a reconstruction of the emission measure distribution along the line of sight. In particular, we select a few locations in the brightest areas of the first brightening region and we use all AIA channels for the reconstruction. 

In order to derive information on the thermal evolution of the observed transient coronal loops, we apply an inversion method \citep{Cheung2015a} to the timeseries of the 6 coronal AIA passband (94~\AA, 131~\AA, 171~\AA, 193~\AA, 211~\AA, 335~\AA) to obtain the differential emission measure. \new{The inversion method considers the full response functions including hot and cool contributions in the double-peaked channels. \cite{Cheung2015a} thoroughly discuss the validation of the method and the estimation of the associated uncertainties; they conclude that the method is overall accurate in recovering the general properties of the DEM and its evolution (though more detailed properties, such as, e.g., DEM slopes at the high temperature end might not be constrained too well).}

\begin{figure}              %%%%%%Figura%%%%%%%%%
 \centering
 \subfigure[]
   {\includegraphics[width=7cm]{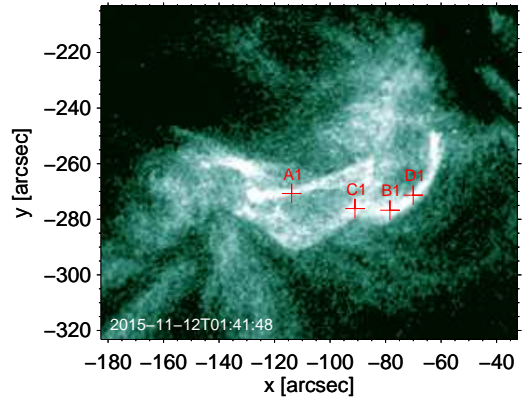}}
 \subfigure[]
   {\includegraphics[width=12cm]{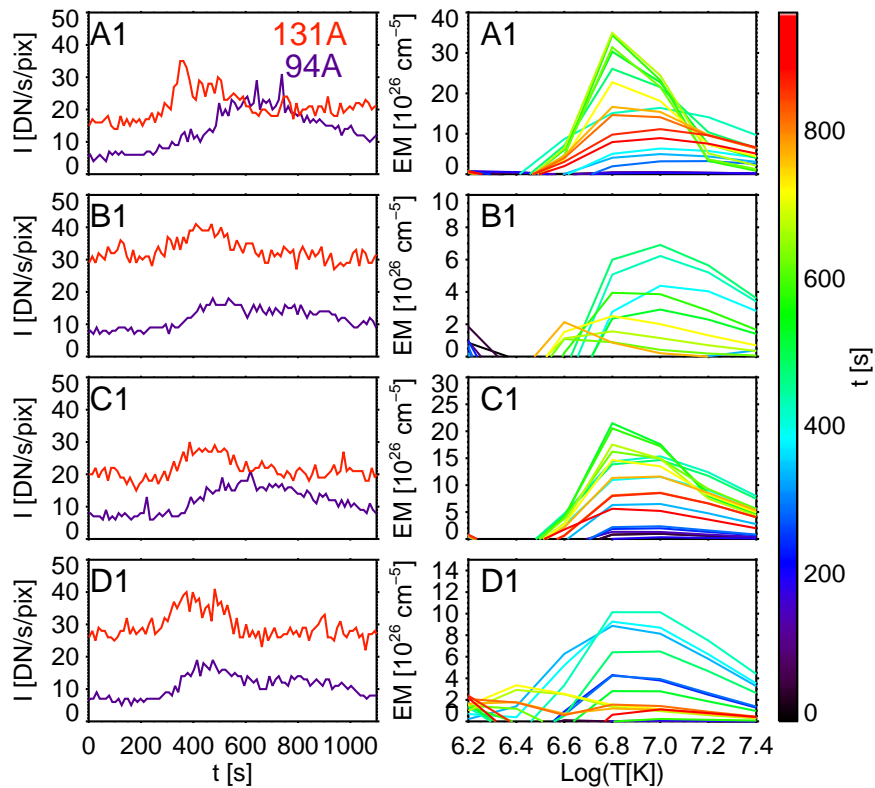}}
\caption{\footnotesize Light curves in the AIA 94\AA\ and 131\AA\ passbands (b, left column) and evolution of emission measure distribution (b, right column), for four locations (marked in panel a) in the hot transient coronal loops shown in Figs.~\ref{fig:reg1_fov}-\ref{fig:reg1_lc} (Event~9 in Table~\ref{table1}). The emission measure distributions vs temperature are shown at progressive times from 0 (black) to 1000s (red). For both lightcurves and EM vs T the reference time used is 01:33:54~UT.} 
\label{fig:em}
\end{figure}

We subtract as background an image for each passband taken before the events starts (2015-11-12T01:37:54).
The observed emission ($I_{i}$, in units of ${\rm DN} s^{-1} {\rm pix}{^-1}$) in each of these AIA narrow-band EUV channels depends on the thermal properties of the optically thin coronal plasma in the pixel, as:
\begin{equation}
I_{i} = \int_{T} R_{i}(T) DEM(T) \,dT
\label{eq:i_dem}
\end{equation}
where $R_{i}(T)$ is the response function in a given passband (in units of ${\rm DN} {\rm cm}^5 s^{-1} {\rm pix}^{-1})$, and the differential emission measure (in units of ${\rm cm}^{-5} {\rm K}^{-1}$) is defined by $DEM(T) \,dT = \int_{z} n_e^2(T) \,dz$, where $n_e^2(T)$ is the electron density of the plasma at temperature T.
We will show plots (Fig.~\ref{fig:em}) of the distribution of emission measure (EM, in units of ${\rm cm}^{-5}$) as a function of temperature, which is obtained by integrating the $DEM(T)$ over 0.2 $\log T[K]$ temperature ranges.

\begin{figure}              %%%%%%Figura%%%%%%%%%
 \centering
   {\includegraphics[width=16cm]{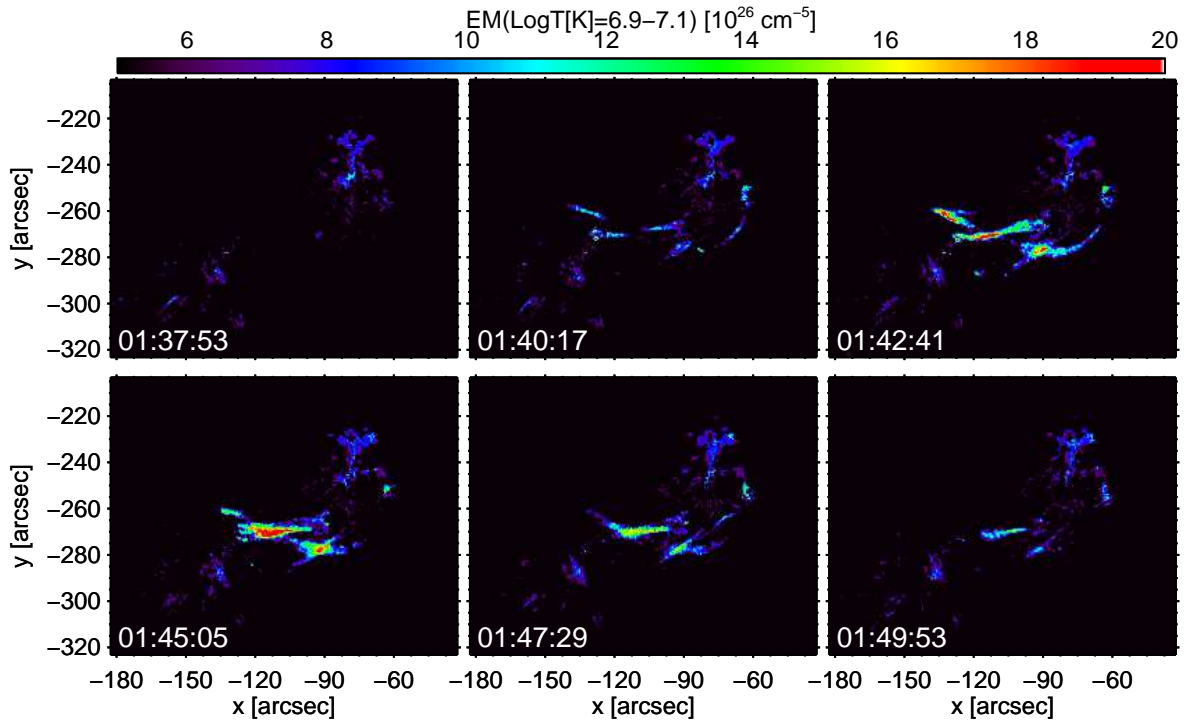}}
\caption{\footnotesize Maps of emission measure in the temperature bin $\log T[K] = 6.9-7.1$ at the 6 labelled times for the same event as in Fig.~\ref{fig:em}. } 
\label{fig:hot_em}
\end{figure}

%The selected locations and the results of the reconstruction are shown in Fig.~\ref{fig:em}. 
The light curves are taken in single pixels and are therefore noisier than those shown in Fig.~\ref{fig:reg1_lc}, but each of them equally shows a well-defined peak that is delayed in the 94~\AA\ channel (Fig.~\ref{fig:reg1_lc}b). \new{The peaks occur between 400~s and 600~s and the delay of the 94~\AA\ channel looks somewhat larger in points A1 and C1, possibly connected to the fact that they are at the intersection of different structures. } The distributions in the right column of Fig.~\ref{fig:em} have clear peaks generally around  $\log T[K] \sim 6.8$. However, a very hot and significant component above $10^7$~K is present, thus confirming the evidence from simple inspection of the images and from light curves. \new{Also in this case, the evolution of the distributions in A1 and C1 is similar and looks different from that in B1 and C1. In A1 and C1 we see a smooth increasing and decreasing trend and the peak seems to be hotter when the emission measure is lower. In B1 and D1 the distributions are generally broader, do not shift in temperature, and their evolution is less smooth. The spatial and temporal coherence of the DEMs, and in particular of its hot components, also supports the reliability of the DEM results obtained with the method (see Fig.~\ref{fig:hot_em}).}

%[20151112_011950]-94:[18,22,38,60]
%[20151112_011950]-131: [18,22]

%[2014/09/18, 20140918_080253] 131:[25(24)]+31
%[2014/09/18, 20140918_080253] 94:[25,28]+31

%[2014/09/17]

\section{Discussion} 
\label{sec:discus}

The systematic analysis of the coronal loops overlying the brightenings imaged by the IRIS mission, especially \neww{in the Si\,{\sc iv} 1400~\AA\ passband} shows that in all cases they correspond to the ignition of complex loop systems to high temperatures (8 - 10~MK) very bright in the hottest SDO/AIA EUV channels, namely the 94~\AA\ and the 131~\AA\ channels.

One important implication of the observed evolution in the hot 94~\AA\ and 131~\AA\ AIA channels and of the emission measure analysis is that the involved plasma is heated above typical persistent \new{active} region temperatures, with \new{detected peak temperatures typically up to $8 - 10$~MK. We note that \cite{Mitra-Kraev2019a} find a case of a microflare, in which the 94~\AA\ emission can be explained by plasma at lower temperatures ($\sim 5$~MK), in contrast with these findings.}

The timescales involved are longer than expected. The loop decay times can be estimated as:

\begin{equation}
\tau_d \approx 15 \frac{L_{Mm}}{\sqrt{T_7}} ~~~~ s
\label{eq:taud}
\end{equation}
where $L_{Mm}$ is the loop half length in units of Mm and $T_7$ is the loop maximum temperature in units of $10^7$ K \citep{Reale2007b}. If we consider the case shown in Figs.~\ref{fig:reg1_94}-\ref{fig:reg1_lc}, for the maximum possible half-length ($\sim 50$~Mm and for a maximum temperature of $\sim 10$~MK (the dependence is relatively weak), we obtain $\tau_d \sim 800$~s. This might look in relative agreement with the observed decay time in Fig.~\ref{fig:reg1_lc}, but we should consider that the channels are narrow-band filters, \new{with relatively narrow ranges of temperature sensitivity, typically much narrower than broad-band X-ray filters \citep[e.g.,][]{Narukage2011a}}, and the emission should be observed for a much shorter time. As a rough estimate, assuming an exponential cooling, the time interval $\Delta t$ in which the emission is observed in a given channel is:

\begin{equation}
\Delta t \approx -\tau_d \ln \left( 1 - \frac{\Delta T}{T_0} \right)
\end{equation}
where $\Delta T$ is the temperature range which the channel is sensitive to, $T_0$ is the maximum temperature. Taking  $\Delta T \sim 5$~MK (this should hold for both channels) and $T_0 \sim 15$~MK as typical values, we obtain $\Delta t \sim 250$~s, much less than the time we observe emission in the hot channels. There are at least two good reasons why we observe the emission for longer than expected. One is that Eq.(\ref{eq:taud}) was derived for conditions of energy equilibrium\citep{Serio1991a}. Energy equilibrium implies that the density is very high to match the high maximum temperature. This is probably not the case here, because the evolution suggests a short heat pulse which does not allow each loop to reach such high density. A lower density drives a slower radiative cooling. The second reason is that, as apparent from the observation itself, we are not capturing a single heating episode but most probably several of them, which determine the brightening of several loops. The light curve is therefore the envelope of unresolved brightenings, and we are more likely detecting the evolution of the total energy release in the loop system. The same kind of analysis can be applied to the other events, and it is even more evident for the third where we are clearly detecting two distinct heating episodes.

As a general overview, all the coronal loop heating events for which IRIS \new{observes} rapid footpoints brightenings in the transition region and chromosphere share common features, and in particular a complex magnetic configuration in the corona. We see in all cases misaligned magnetic arches that most likely interact with each other determining impulsive energy releases. We may simply address them as large scale magnetic rearrangements, probably driven by large scale photospheric motions or large scale magnetic flux emergence. The interaction of misaligned magnetic channels determines large scale magnetic reconnection, which in turn leads to an impulsive energy release coherent in space and time across the structures. Independent signatures of this are the coherent brightening in the hottest AIA channels, sensitive to emission from $\log T[K] \gtrsim 6.7$ plasma and the presence of particle acceleration as obtained from hydrodynamic modeling \citep{Testa2014a,Polito2018a}. \new{\cite{Ugarte-Urra2014a, Ugarte-Urra2017a, Ugarte-Urra2019a} have studied transient loops in active region cores observed in the Fe\,{\sc xviii} emission of the AIA 94~\AA\ passband, and investigated: (a) their frequency \citep{Ugarte-Urra2014a}, (b) the relation between total Fe\,{\sc xviii} emission and total unsigned flux of the active region \citep{Ugarte-Urra2017a}, and (c) the relation between the Fe\,{\sc xviii} emission in each of these impulsively heated loops and the magnetic ($B_{avg}$) and geometric (loop length, $L$) properties of the loop \citep{Ugarte-Urra2019a}.
Here we focus on the observed morphology of the AIA 94~\AA\ and 131~\AA\ hot emission in each of these IRIS events, which suggests that the impulsive heating derives from large scale magnetic rearrangements/large angle reconnection. 
The interaction of small loops has been taken as explanation of a small flare observed with IRIS \citep{Alissandrakis2017a}, and our study provides direct evidence from the morphology observed in the hot AIA channels counterparts.
Many recent papers have studied simultaneous observations from IRIS and SDO/AIA to investigate magnetic rearrangements causing localized brightenings in the transition region, and likely related with flux emergence \citep[e.g.,][]{Jiang2015a,Toriumi2017a,Guglielmino2018a,Huang2018a,Tian2018a,Guglielmino2019a}. However we note that the coronal heating events we study here have quite different properties with respect to those events, in that they have generally lower Doppler shift velocities (up to $\sim \pm 30$~km/s) and show heating of the overlying coronal loops to several MK. }

The evolution and the energetic scale of the events make them intermediate -- and are therefore valuable links -- between intense flares and diffuse nanoflare activity. They are also at the right scale to be as intense and coherent as to let us study their evolution in good detail.

The observed complex evolution cannot be described by simple hydrodynamic models and requires detailed MHD modeling. Although self-standing, a natural extension of this work will be the attempt to model this specific kind of large-scale magnetic interaction through detailed MHD simulations, which will be the subject of a forth-coming work.
It will also be very important to acquire more complete observational evidence, through large band spectroscopy.

%- Images and evolution show large scale magnetic interaction and rearrangement

%- Hot loop ignition due to interacting magnetic tubes 

%- Episodes with severe and large-scale magnetic rearrangement that determines fast reconnection, particle acceleration and impulsive heating

%- Intermediate events between nanoflares and major flares

%- Modeling confirms that resistive reconnection is a necessary ingredient to have loop ignition in the hot AIA channels

\acknowledgments{F.R., A.P., acknowledge support from Italian Ministero dell'Istruzione, dell'Universit\`a e della Ricerca. PT acknowledges support by NASA grants NNX15AF50G and NNX15AF47G, and by contracts 8100002705 and SP02H1701R from Lockheed-Martin to SAO. The authors thank the International Space Science Institute (ISSI) for their support and hospitality during the meetings of the ISSI team “New Diagnostics of Particle Acceleration in Solar Coronal Nanoflares from Chromospheric Observations and Modeling.”}

\appendix
\section{Other events}

\begin{figure}              %%%%%%Figura%%%%%%%%%
 \centering
    \subfigure[]
   {\includegraphics[width=6cm]{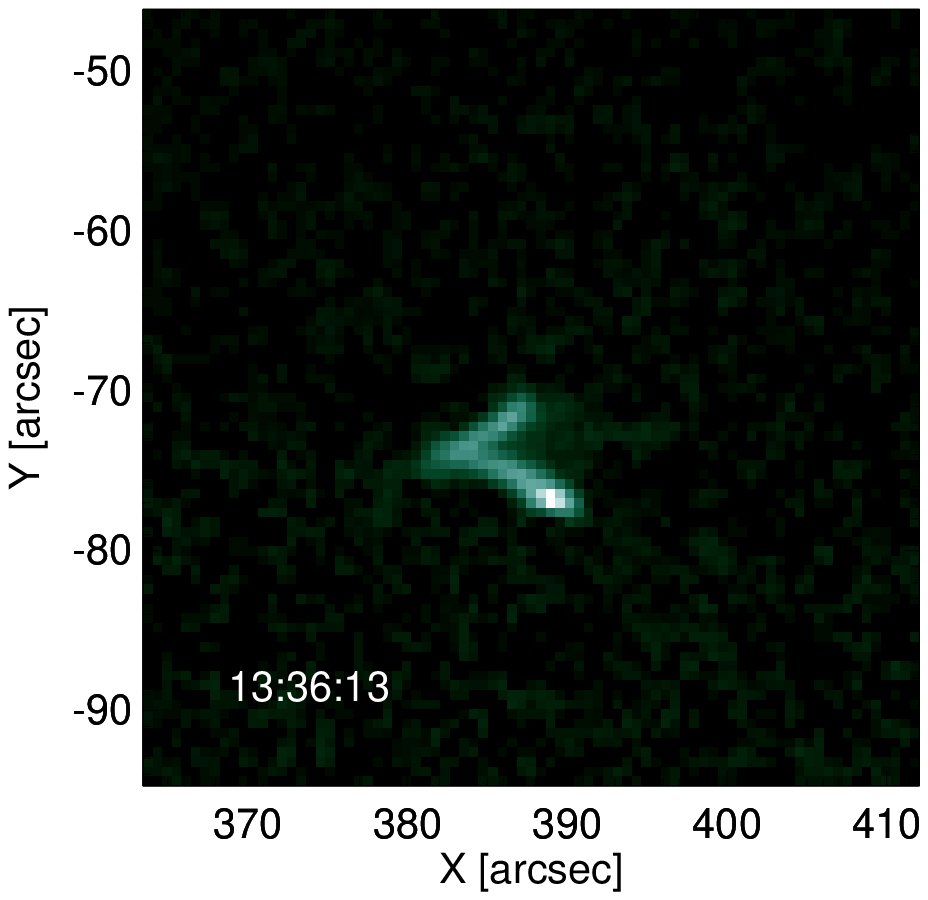}}
 	\subfigure[]
   {\includegraphics[width=6cm]{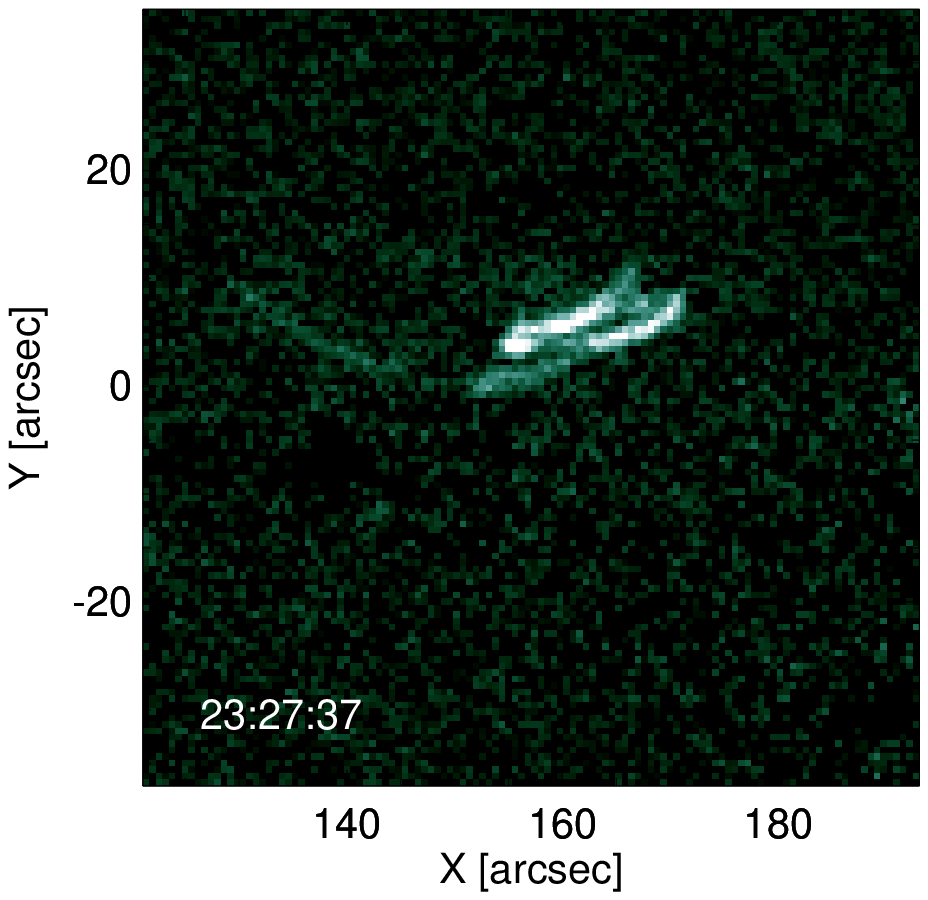}}
	\subfigure[]
   {\includegraphics[width=6cm]{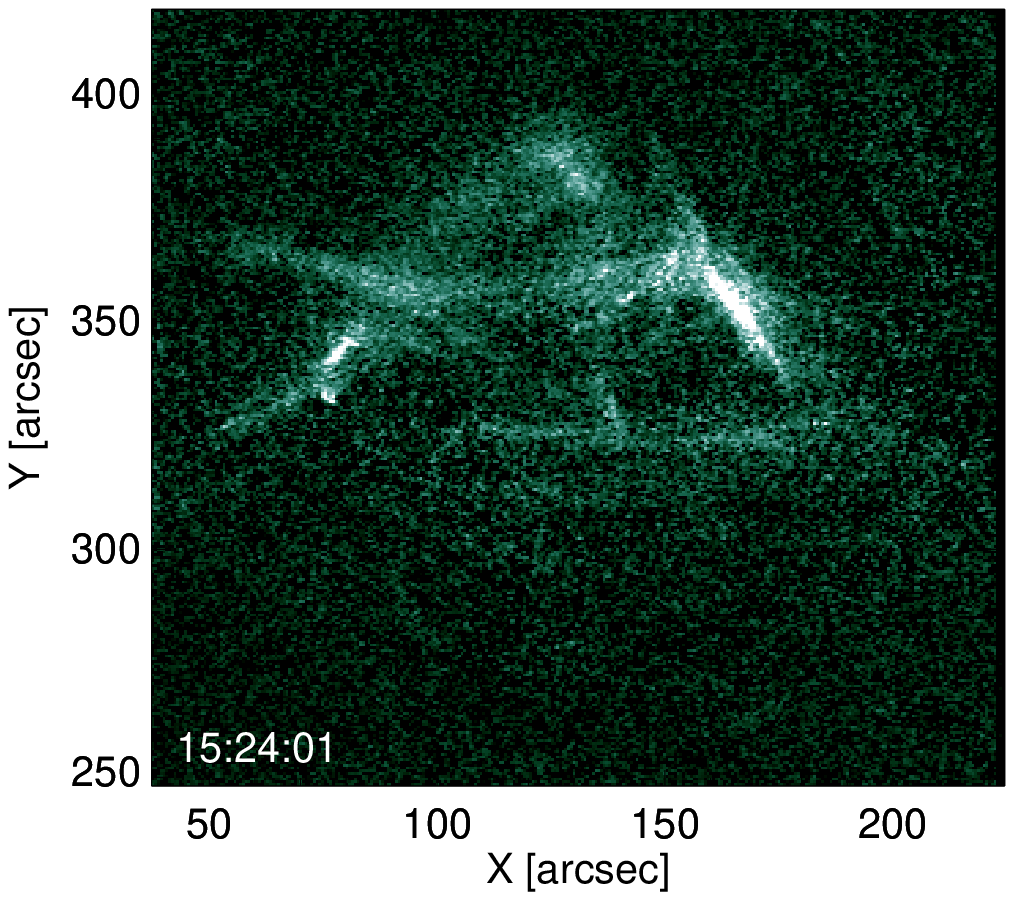}}
    \subfigure[]
   {\includegraphics[width=6cm]{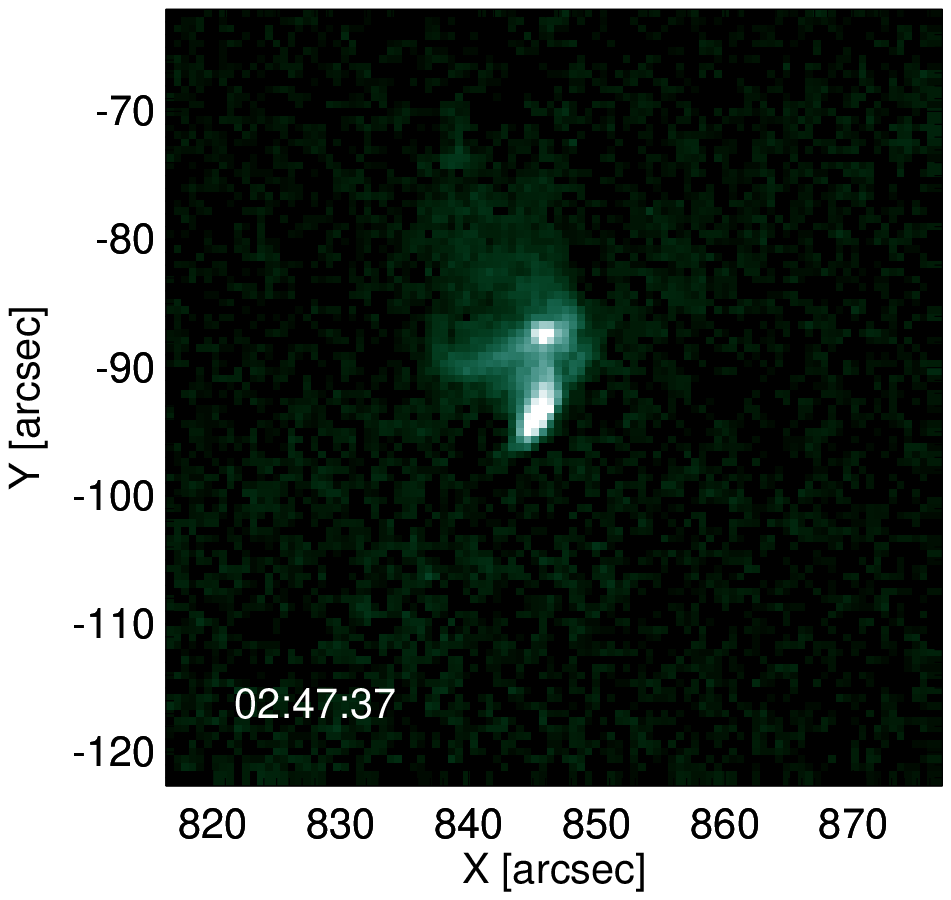}}
    \subfigure[]
   {\includegraphics[width=6cm]{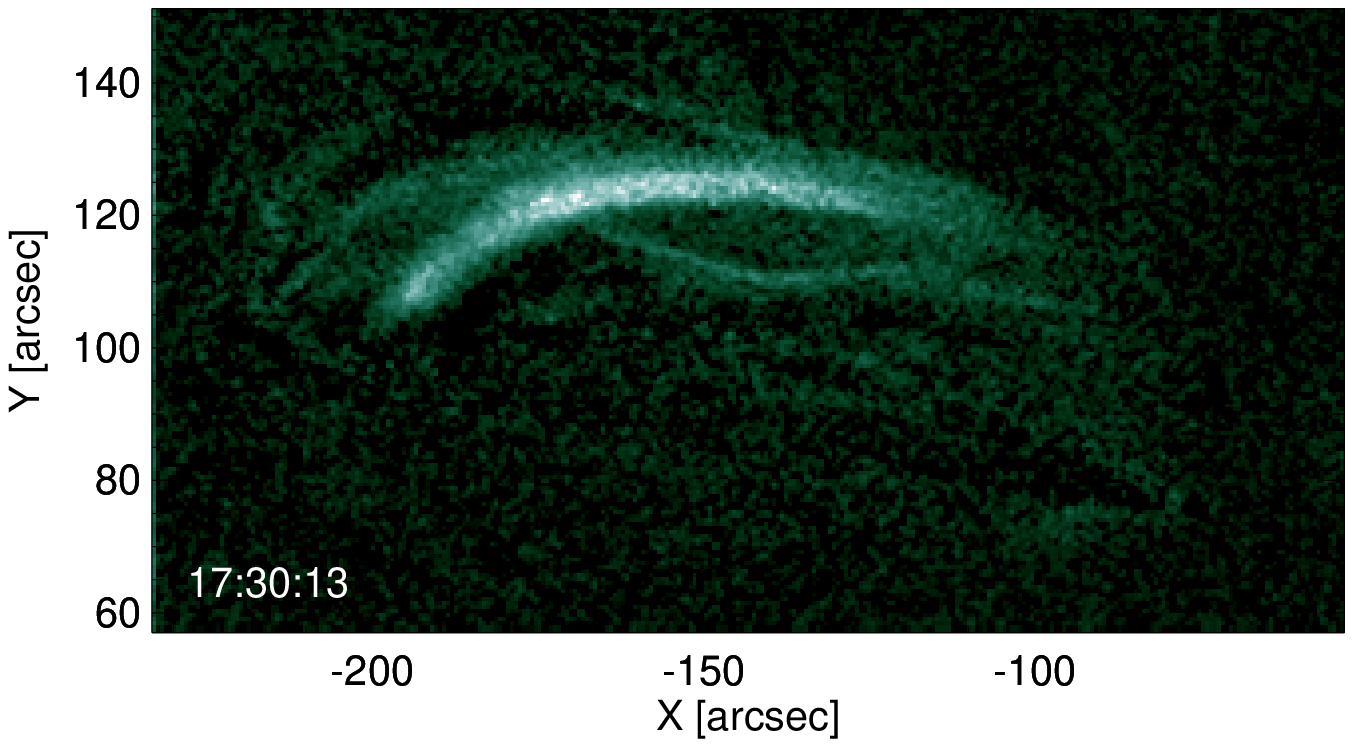}}
    \subfigure[]
   {\includegraphics[width=6cm]{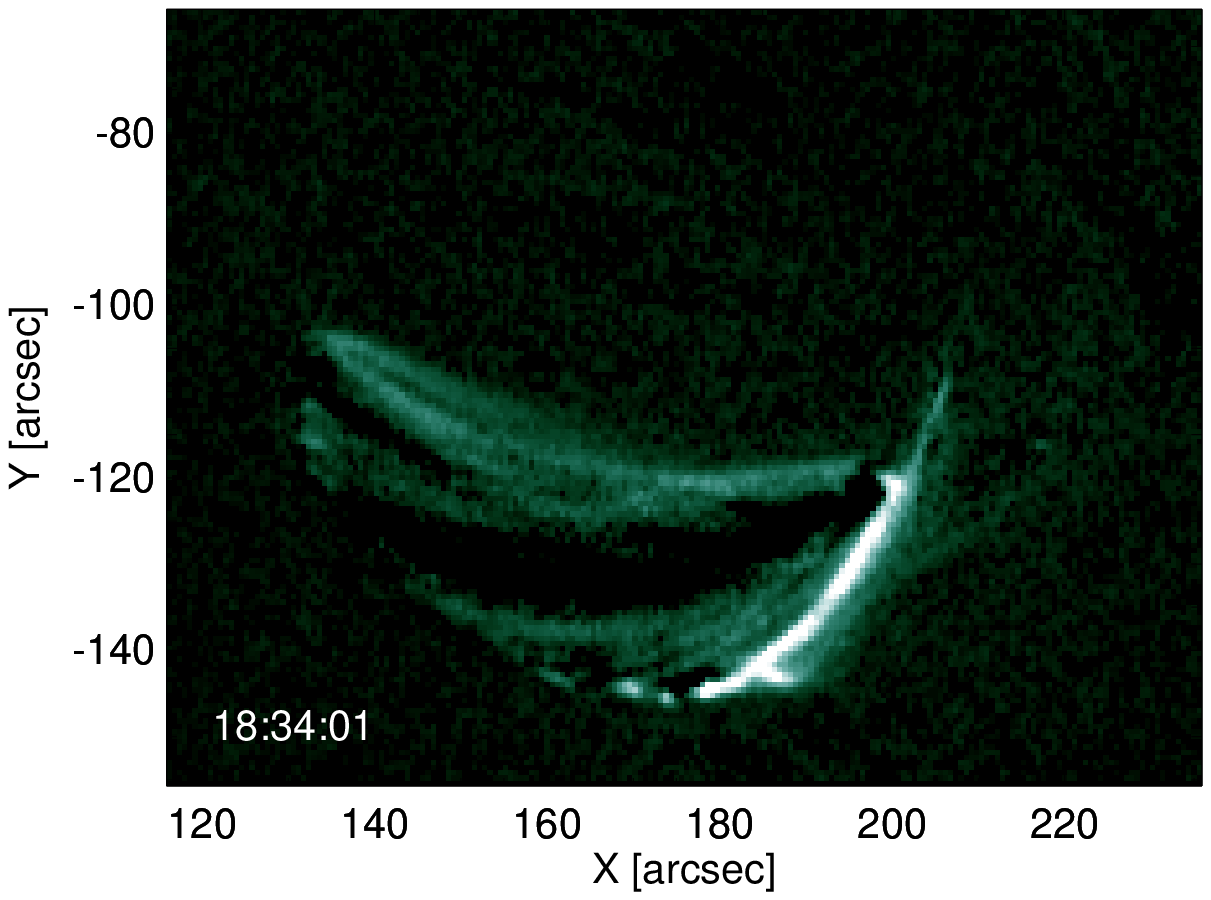}}      
    \subfigure[]
   {\includegraphics[width=6cm]{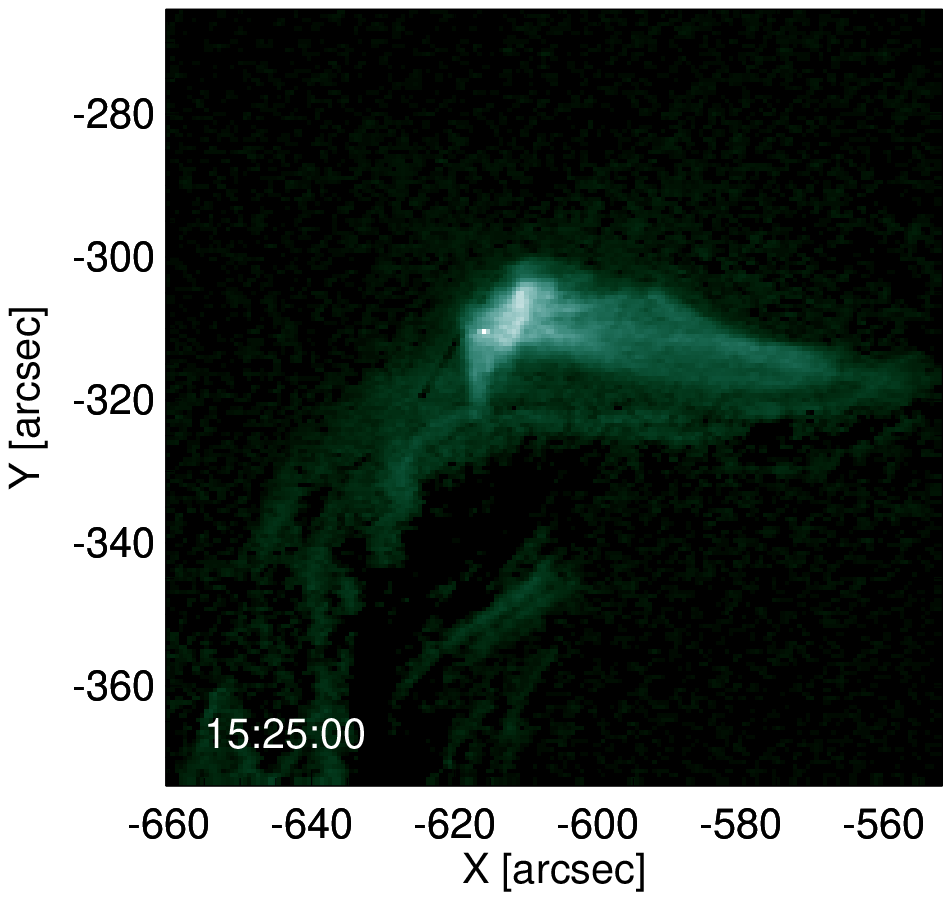}}
\caption{\footnotesize Representative images of the other events (at the labelled times) listed in Table~\ref{table1} taken in the 94~\AA\ channel: (a) \# 1 (2014-02-04), (b) 2 (2014-02-23), (c) 3 (2014-03-19), (d) 4 (2014-04-10), (e) 6 (2014-09-17), (f) 8 (2015-01-29), (g) 10 (2015-12-24). } 
\label{fig:other}
\end{figure}

Fig.~\ref{fig:other} shows representative images of the other events listed in Table~\ref{table1}, taken in the 94~\AA\ channel. All of them show complex magnetic configurations.

\end{document}